\documentclass{aa}  

\usepackage{comment}
\usepackage{tabularx}
\newcolumntype{Y}{>{\centering\arraybackslash}X}
\usepackage{subfigure}
\usepackage{graphicx}
\usepackage{txfonts}
\usepackage[percent]{overpic}
\usepackage[normalem]{ulem}
\usepackage{bm}
\usepackage[usenames,dvipsnames,svgnames,table]{xcolor}
\usepackage[colorlinks=true,linkcolor=blue,citecolor=blue]{hyperref}
\begin{document} 

	\title{Accretion disk coronae of Intermediate Polar Cataclysmic Variables}

	\subtitle{3D~MagnetoHydro-Dynamic modeling and thermal X-ray emission}

	\author{
		E. Barbera\inst{1}\fnmsep\thanks{email: ebarbera@astropa.unipa.it},
		S. Orlando\inst{2}\and
		G. Peres\inst{1,2}
	}

	\institute{
		Dip. di Fisica e Chimica, Specola Universitaria, Universit\`a degli Studi di Palermo, Piazza del Parlamento 1, 90134 Palermo, Italy
		\and
		INAF - Osservatorio Astronomico di Palermo, Piazza del Parlamento 1, 90134 Palermo, Italy\\
	}

	\date{Received <Month> <day>, <year>; accepted <Month> <day>, <year>}

 
	\abstract
	{\emph{Intermediate Polar Cataclysmic Variables} (\emph{IPCV}) contain a magnetic, rotating white dwarf surrounded by a magnetically truncated accretion disk. To explain their strong flickering X-ray emission, accretion has been successfully taken into account. Nevertheless, observations suggest that accretion phenomena could not be the only process behind it. An intense flaring activity occurring on the surface of the disk may generate a corona, contribute to the thermal X-ray emission and influence the system stability.}
	{Our purposes are: investigating the formation of an extended corona above the accretion disk, due to an intense flaring activity occurring on the disk surface; studying its effects on the disk and stellar magnetosphere; assessing its contribution to the observed thermal X-ray flux.}
	{We have developed a 3D magnetohydrodynamic (MHD) model of a \emph{IPCV} system. The model takes into account gravity, disk viscosity, thermal conduction, radiative losses and coronal flare heating through heat injection at randomly chosen locations on disk surface. To perform a parameter space exploration, several system conditions have been considered, with different magnetic field intensity and disk density values. From the results of the evolution of the model, we have synthesized the thermal X-ray emission.}
	{The simulations show the formation of an extended corona, linking disk and star. The flaring activity is capable of strongly influencing the disk configuration and possibly its stability, effectively deforming the magnetic field lines. Hot plasma evaporation phenomena occur in the layer immediately above the disk. The flaring activity gives rise to a thermal X-ray emission in both the $[0.1-2.0]\;keV$ and the $[2.0-10]\;keV$ X-ray bands.}
	{An intense coronal activity occurring on the disk surface of an \emph{IPCV} can affect the structure of the disk depending noticeably on the density of the disk and the magnetic field of the central object. Moreover, the synthesis of the thermal X-ray fluxes shows that this flaring activity may contribute to the observed flickering thermal X-ray emission.}

	\keywords{
		Stars: novae, cataclysmic variables -- 
		Stars: flare -- 
		Magnetohydrodynamics (MHD) -- 
		Accretion, accretion discs -- 
		Stars: coronae -- 
		X-rays: stars  
	}

   \maketitle


\section{Introduction}
\label{sec:Intro}

\emph{DQ Her} stars, also known as \emph{Intermediate Polar Cataclysmic Variables} (\emph{IPCV}s), contain an accreting, magnetic, rapidly rotating white dwarf. The white dwarf in \emph{IPCV}s typically has a magnetic field strength ranging between 0.1 and 10 MG \citep{Hellier2007}, weaker than the field of other classes of \emph{CV}s (e.g. \emph{AM Her} stars also known as \emph{Polar Cataclysmic Variables}). 
In particular, differently from \emph{AM Her} stars, the field is not strong enough to prevent an accretion disk from forming, but it is capable of disrupting the inner part of the disk where the magnetosphere dominates the accretion.
So, in \emph{IPCV}s, accreting matter follows the magnetic field lines within the magnetosphere delimited by the truncation radius $R_{d}$, i.e. where the magnetic pressure is approximately equal to the total gas pressure.

\emph{IPCV}s, with a medium-strength field, combine the characteristics of a non-magnetic system (in its outer regions) with those of a strongly magnetized object (in the region close to the white dwarf). 
Moreover, \emph{IPCV} class are also strong X-ray emitters and the ratio between X-ray and visible flux ($F_{x}/F_{v}$) is greater than unity for this class of stars \citep{Patterson1994}. These stars are indeed soft X-ray $[0.1-2.0]\;keV$ and hard X-ray $[2.0-10]\;keV$ sources with asynchronous changes of brightness.
Nevertheless, while X-rays from \emph{IPCV}s do exhibit aperiodic variability, this is not the most notable aspect of the X-ray timing properties of these objects. In fact, one of the defining characteristics of \emph{IPCV}s is the strictly periodic and coherent X-ray spin modulation, i.e. the presence of a rapid and stable periodicity in the light curve of the \emph{CV} typically observed at optical or X-ray wavelengths with $P_{spin}<P_{orb}$, although most stars have $P_{spin}\ll P_{orb}$. 
There is a large consensus in the literature that the origin for this X-ray emission is the mass accretion \citep{Patterson1994}.
The value of mass accretion rate $\dot{M}$ in \emph{IPCV}s has been inferred from observations through various techniques based on different assumptions, and the expected value of $\dot{M}$ usually ranges between $10^{-11}$ and $10^{-8}\; M_{\odot}y^{-1}$ \citep{Patterson1984,Patterson1994,Giovannelli2012}. 
A significant amount of energy per particle is therefore released, due to shocks generated by the accreting plasma flowing along the field lines and impacting onto the white dwarf's surface at, approximately, free fall velocity.
According to the model of accreting white dwarf, these impacts should cause a strong emission in the hard X-ray band $[2-10\;keV]$ up to $L_{hx}\approx10^{34}\;erg\;s^{-1}$ \citep[e.g.][]{Patterson1994}.

Another intriguing feature shared by all the \emph{CV} stars, including \emph{IPCV}s, is the \emph{flickering}, i.e. a random, broad-band and short term modulation of luminosity. It appears as a sequence of overlapping flares and bursts apparently having no cycle or periodicity.
This phenomenon has been observed in the X-ray light curves of \emph{CV}s of all classes, and ranges from fluctuations lasting few seconds to larger changes in brightness with rises and dips lasting for hours, with no pattern or period.
The fact that \emph{flickering} is a characteristic feature of all \emph{CV}s immediately suggests that more than one mechanism must be involved. This is supported by the statistical properties of the \emph{flickering} of the various type of \emph{CV} \citep{Bruch2000}.	
In fact, studies of the \emph{flickering} in \emph{CV} stars conducted by \cite{FritzBruch1998} suggest an origin of this phenomenon for magnetized \emph{CV}s (such as \emph{IPCV}s) leading to properties which are distinct from those of non-magnetic \emph{CV}s. In particular, \emph{IPCV}s show a strong \emph{flickering} on short time-scales with a high strength.
Observations suggest that in some systems \emph{flickering} may arise in the turbulent inner region of the disk, or from the bright spots on the surface of the white dwarf \citep{Hellier2001book}. 
Actually, the analysis of eclipsing lightcurves \citep{Horne-Stiening1985} demonstrated that the \emph{flickering} source must be concentrated in the inner portion of the accretion disk but it is not necessarily confined to the immediate vicinity of the white dwarf, possibly spreading out in the inner part of the accretion disk \citep{Bruch2015}.

The differentially rotating disk is thought to be a turbulent environment. Theoretical studies of \cite{Balbus&Hawley_I-II_1991} predicted that the turbulence in accretion disks may be driven by \emph{Magneto-Rotational Instability} (\emph{MRI}) and this phenomenon has also been observed through numerical models \citep[e.g.][]{Hawley1991,Stone1996}. 
This may lead to magnetic reconnection phenomena. 
To this regard we should note that significant parts of the disk can be turbulent, therefore, reconnection is possible at different distances from the star.
On the other hand, the studies of \cite{GaleevRosnerVaiana1979} showed that the differential rotation of the Keplerian disk is able to lead to the winding of the magnetic field lines, increasing the field and determining the phenomenon of the expulsion of the magnetic field from the disk in the form of loops. Through the magnetic energy release, due to magnetic reconnection, an extended magnetic corona can be built up and sustained \citep{Beloborodov1999,MalzacBeloborodovPoutanen2001}. 
Similar phenomenon has been observed in simulations of \emph{MRI} disks, where the magnetic corona has been formed and expanded above and below the disk due to winding of the field lines and amplification of the magnetic field \citep{Armitage2002,Steinacker&Papaloizou2002}. 
The formation of extended corona has also been observed in simulations of accretion onto a magnetized stars from MRI-driven disk \citep{Romanova2011}.
Hence, the combination of a turbulent disk and its differential rotation may trigger an exponential amplification of a seed magnetic field \citep{Beloborodov1999,GaleevRosnerVaiana1979}. The resulting magnetic field is sufficiently strong to cause magnetic reconnection close to the disk surface where the observations suggest that phenomena responsible of \emph{flickering} may occur \citep{Hellier2001book,Seward-Charles2010book}.

In the light of these considerations, it is plausible that the X-ray emission observed in some of the \emph{IPCV} stars may have a thermal component due to a flaring activity occurring on the surface of the accretion disk. This flaring activity is to some extent similar to the flares taking place in the corona of the Sun and solar-like stars, and may develop and sustain an extended corona. 

It is worth noting that in the models cited above, the presence of the stellar magnetic field is not required for the formation of an extended corona in the disk; in fact, high-energy X-ray coronae are also observed in objects, where the stellar magnetic field is not present (e.g. black holes).
In particular for \emph{CVs}, there are no remarkable differences in the properties of extended coronae observed from magnetic (such as \emph{IPCV}s) and non-magnetic \emph{CVs}, since the main features differentiating these two types of \emph{CVs} are: the presence of circular polarization of the emission due to the interaction of accreting charged material with magnetic field; 
a spin period lower than orbital period $P_{spin}\ll P_{orb}$; 
$He\;II$ emission lines due to ionization of the accretion columns by the EUV continuum of the radial accretion shock \citep{Szkody2003}; 
and the properties of \emph{flickering} \citep{FritzBruch1998}, whose origin may be mainly related to accretion, that in magnetized objects is thought to occur via accretion columns or curtains \citep{Romanova2002,Romanova2003}.

Here we perform 3D MHD simulations of the formation of accretion disk coronae in different prototypes of \emph{IPCV}s through a hail of flares occurring on the surface of the disk. The flares are stocastically triggered injecting energy in the system by means of a heat pulse with random intensity. We consider a magnetized central object with different intensities of the magnetic field. Although this is not a requirement for the formation of the corona, in this way we may reproduce the formation of coronal loops linking star and disk. Similar structures have been predicted using MHD models \citep{Orlando2011} and observed in young stellar objects \citep{Favata2005}.

\section{Model and numerical setup}
\label{sec:Model}


The model describes an intense flaring activity taking place on the surface of the accretion disk placed around a rotating magnetized compact object, a prototype of \emph{IPCV}. 
The central compact object is surrounded by a structured atmosphere initially unperturbed and approximately in equilibrium. In order to study \emph{IPCV}s, we adapted the model described in \cite{Orlando2011} developed for \emph{Classical T-Tauri Stars} (\emph{CTTS}).


\subsection{MHD equations}
The system evolution can be described through the four time-dependent classical MHD equations, extended to include gravitational force, viscosity, thermal conduction, radiative cooling and plasma heating. These equations are defined in a 3D spherical coordinate system ($r$, $\theta$, $\phi$) and they are here reported in c.g.s. in the conservative form:

\begin{equation}
	\frac{\partial \rho}{\partial t}+ \nabla\cdot\left( \rho \vec{ u } \right) = 0 
	\label{eq:MassCon}
\end{equation}
\smallskip
\begin{equation}
	\frac{\partial \rho\vec{ u }}{\partial t}+ \nabla \cdot \left( \rho \vec{ u }\vec{ u } \right) = 
	-\nabla p
	+\left( \frac{\vec{j}\times\vec{B}}{c} \right) 
	+\nabla\cdot\vec{\tau}
	+\rho \vec{g}
	\label{eq:MomCon}
\end{equation}
\smallskip
\begin{equation}
	\begin{split}
	\frac{\partial \varepsilon}{\partial t}+ \nabla \cdot \left( \varepsilon \vec{u} \right) =
	&-\nabla\cdot\left(p\vec{u} \right)  
	+\vec{u}\cdot\left(\frac{\vec{j}\times\vec{B}}{c}\right)
	+\nabla\cdot\left(\vec{u}\cdot\vec{\tau}\right)
	+\rho\vec{u}\cdot\vec{g}\\
	&-\nabla\cdot\vec{F_{c}} 
	-n_{e}n_{H}\Lambda\left( T \right) 
	+Q\left( r,\theta,\phi,t \right) 
	\end{split}
	\label{eq:EneCon}
\end{equation}
\smallskip
\begin{equation}
	\frac{\partial \vec{B}}{\partial t}=-c\nabla\times\vec{E}=\nabla\times\left(\vec{u}\times\vec{B}\right)
	\label{eq:InductionEq}
\end{equation}
\\
where $p$ is the thermal pressure for a fully ionized plasma, and the total energy per unit volume ($\varepsilon$) is the sum of internal or thermal ($\varepsilon_{i}$) and kinetic energy ($E_{k}$), with $\gamma=5/3$:

\begin{equation}
	p = 2n_{H}k_{B}T; 
	\qquad\qquad
	\varepsilon = \varepsilon_{i} + E_{k} = 
	\frac{p}{\left(\gamma - 1\right)} + \rho\frac{\vec{u}\cdot\vec{u}}{2}.
	\label{eq:TotPress_TotEnegy}
\end{equation}
\\
The mass density is defined as $\rho=\mu m_{H} n_{H}$, where $m_{H}$ is the mass of hydrogen atom, $n_{H}$ is the number of hydrogen atoms per unit of volume, and $\mu=1.28$ (assuming metal abundances of 1, equal to solar ones) is the mean atomic mass \citep{AndersGrevesse1989}. The bulk velocity of the plasma is defined as $\vec{u}$.
Moreover, the Eq. (\ref{eq:InductionEq}) has been taken into account in the ideal limit of high \emph{magnetic Reynolds number} $(R_{m}\gg1)$.

The viscosity is taken into account in MHD equations through the viscous tensor $\vec{\tau}$. The viscosity is assumed to be effective only in the circumstellar disk and negligible in the region of the extended corona \citep{Orlando2011,Giovannelli2012}. The technique used to track the disk material, where viscosity is effective, is based on the definition of a tracer following the same evolution of density (i.e. the tracer is passively advected similarly to the density). 
We defined the tracer $C_{d}$ as the disk mass fraction inside each computational cell, initializing the disk material with $C_{d}=1$ and the extended corona with $C_{d}=0$.
Then the viscosity is enabled only in the computational cells containing at least 99\% of material of the disk (i.e. $C_{d}>0.99$). The viscous stress tensor is defined as:

\begin{equation}
	\vec{\tau} = \eta_{v}\left[
	\left( \nabla \vec{u} \right) +
	\left( \nabla \vec{u} \right)^{T} - 
	\frac{2}{3} \left( \nabla \cdot \vec{u} \right) \vec{I} 
	\right] 
	\label{eq:ViscTensor}
\end{equation}
\\
where dynamic (or shear) viscosity $\eta_{v}$ is proportional to kinematic viscosity $\nu_{v}$ and density as follows:

\begin{equation}
	\eta_{v} = \nu_{v} \rho = \rho\alpha \frac{c_{s}^{2}}{\Omega_{k}}
	\label{eq:DynVisc}
\end{equation}
\\
where $c_{s}=\sqrt{\gamma p / \rho}$ is the isothermal sound speed and $\Omega_{k}$ is the angular velocity of the circumstellar disk in quasi-Keplerian rotation. $\alpha$ is a phenomenological dimensionless parameter defined in the Shakura \& Sunyaev model taking into account the  effect of anomalous viscosity \citep{ShakuraSunyaev1973}. In this model $\alpha$ can be assumed to be a measure of the strength of the turbulent angular momentum transport and varies in the range $0.01-0.6$ \citep{Balbus2003}. In the simulations here presented, we assumed constant $\alpha = 0.02$ \citep{Romanova2002}.

Thermal conduction appears in the energy Eq. (\ref{eq:EneCon}) as minus the heat flux divergence ($-\nabla \cdot \vec{F_{c}}$). In accordance with the standard formulation by Spitzer \citep{Spitzer1962}, thermal conduction is highly anisotropic due to the presence of the magnetic field that drastically reduces the conductivity component transverse to the field. As a matter of fact, thermal flux can be locally splitted in two components, one along and the other across the magnetic field lines:

\begin{equation}
	\begin{split}
	\mathcal{L}_{c} =
	-\nabla\cdot\vec{F_{c}} = 
	-\nabla_{\parallel} \cdot \vec{F_{\parallel}}  -\nabla_{\perp} \cdot \vec{F_{\perp}}.
	\label{eq:ThermCond}
	\end{split}
\end{equation}
\\
The two components of thermal flux can therefore be written as

\begin{equation}
	\begin{split}
	&F_{\parallel} = \left[ q_{spi} \right]_{\parallel} = -k_{\parallel} \left( \nabla_{\parallel}T \right) \approx -9.2 \cdot 10^{-7} T^{\frac{5}{2}} \left( \nabla_{\parallel}T \right)
	\\ 
	&F_{\perp} = \left[ q_{spi} \right]_{\perp} = -k_{\perp} \left( \nabla_{\perp}T \right) \approx -3.3 \cdot 10^{-16} \frac{n_{H}^{2}}{T^{\frac{1}{2}}B^{2}} \left( \nabla_{\perp}T \right)
	\label{eq:ThermCond_q_spi}
	\end{split}
\end{equation}
\\
where $k_{\parallel}$ and $k_{\perp}$ are the thermal conduction coefficients along and across field lines (in units of $erg s^{-1} K^{-1} cm^{-1}$). From Eq.  (\ref{eq:ThermCond_q_spi}) it is clear that, if the magnetic field is sufficiently strong, the ratio ${k_{\perp}}/{k_{\parallel}} \ll 1$ and thermal conduction behaves anisotropically working prevalently along field lines.
Moreover, the classical theory of thermal conduction is based on the assumption that the mean free path $\lambda$ is smaller than the temperature scale height $L_{T}$, and when the requirement

\begin{equation}
	\lambda \ll L_{T} = \frac{T}{\left| \nabla T \right|} 
	\label{eq:ThermCond_lambda}
\end{equation}
\\
is not fullfilled, the Eq. (\ref{eq:ThermCond}) overestimates thermal flux. 
When fast phenomena occur, like flares in their early phases, it is well known that rapid transients, fast dynamics and steep gradients develop \citep{RealeOrlando2008}. So, under these circumstances, Spitzer's heat conduction formulation breaks down becoming flux limited, and therefore cannot be suitable \citep{Brown1979}.
For this reason the two components of thermal conduction must be redefined as follows \citep{Orlando2008}:

\begin{equation}
	\begin{split}
	&F_{\parallel} = \left( \frac{1}{\left[ q_{spi} \right]_{\parallel}} + \frac{1}{\left[ q_{sat} \right]_{\parallel}} \right)^{-1} 
	\\	
	&F_{\perp} = \left( \frac{1}{\left[ q_{spi} \right]_{\perp}} + \frac{1}{\left[ q_{sat} \right]_{\perp}} \right)^{-1}
	\label{eq:ThermCond_F_spi_sat}
	\end{split}	 
\end{equation}
\\
where the saturated heat flux along and across field lines are \citep{CowieMcKee1977}

\begin{equation}
	\begin{split}
	&\left[ q_{sat} \right]_{\parallel} = -sign \left( \nabla_{\parallel}T \right) \cdot 5\varphi\rho c_{s}^{3}
	\\ 
	&\left[ q_{sat} \right]_{\perp} = -sign \left( \nabla_{\perp}T \right) \cdot 5\varphi\rho c_{s}^{3}
	\label{eq:ThermCond_q_sat}
	\end{split}	 
\end{equation}
\\
$\varphi$ is a parameter we set equal to unity \citep{Giuliani1984,Borkowski1989,Fadeyev2002}.

The calculation were performed using PLUTO \citep{Mignone2007}, a modular Godunov-type code developed mainly for astrophysical plasmas involving high Mach number flows in multiple spatial dimensions. The code embeds different physics modules (allowing the treatment of hydrodynamics and magneto-hydrodynamics both relativistic and non-relativistic in Cartesian or curvilinear coordinates) and multiple algorithms particularly oriented toward the treatment of astrophysical flows in the presence of discontinuities. These features make PLUTO a flexible and versatile modular computational framework to solve the equations describing astrophysical plasma flows. PLUTO is entirely written in the C programming language and was designed to make efficient use of massive parallel computers using the message-passing interface (MPI) library for interprocessor communications. The MHD equations are solved using the available MHD module, configured to compute inter-cell fluxes with the Harten-Lax-Van Leer (HLL) approximate Riemann solver. 
In order to solve the equation vs. time, we used a second order Runge-Kutta (RK) integration scheme. Moreover, a minmod limiter for the primitive variables has been used, in order to avoid spurious oscillations that could otherwise occur due to shocks, discontinuities or sharp changes in the solution domain. 
As for magnetic field, its evolution is computed adopting the constrained transport approach \citep{BalsaraSpicer1999} that maintains the solenoidal condition $(\nabla\cdot B=0)$ at machine accuracy. Moreover, we adopted the \emph{magnetic field-splitting} technique \citep{Tanaka1994,Powell1999,ZanniFerreira2009} by splitting the total magnetic field into a contribution coming from the background stellar magnetic field and a perturbation to this initial field. Then only the latter component is computed numerically. This approach is particularly useful when dealing with low-$\beta$ plasma as it is the case in proximity of the stellar surface \citep{ZanniFerreira2009}. 
PLUTO takes into account radiative losses from optically thin plasma. The radiative losses are computed as

\begin{equation}
	\Lambda_{i}=n_{e}n_{H}\Lambda(T_{i})
	\qquad
	\text{with}
	\qquad
	n_{e}=n_{H}=\frac{\rho}{\mu m_H}
	\label{eq:Lambda}
\end{equation} 
\\
\begin{figure}[tb!]
	\centering
	\includegraphics[width=0.8\columnwidth]{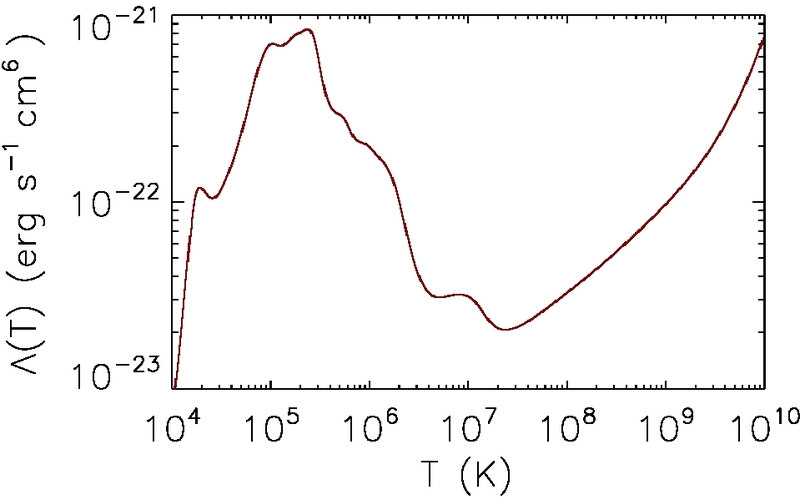}%
	\caption{The radiative losses per unit emission measure $\Lambda(T)$ as a function of temperature.}
	\label{fig:RadCool_cooltable}
\end{figure} 
\\
at the temperature of interest, using a lookup-table/interpolation method\footnote{The lookup table used has been generated with \emph{Cloudy 90.01} for an optically thin plasma and solar abundances, thanks to T. Plewa.}, where $\Lambda_{i}=\Lambda(T_{i})$ is available along with $T_{i}$ as a sampling at discrete point (see Fig. \ref{fig:RadCool_cooltable}). The thermal conduction as well the viscosity is treated separately from advection terms through operator splitting. In particular we adopted the super-time-stepping technique \citep{Alexiades1996} which has been proved to be very effective to speed up explicit time-stepping schemes for parabolic problems. This approach is crucial when high values of plasma temperature are reached (as during flares), explicit scheme being subject to a rather restrictive stability condition (i.e. $\Delta t < (\Delta x)^2/(2\eta)$, where $\eta$ is the maximum diffusion coefficient), as the thermal conduction timescale $\tau_{cond}$ is typically shorter than the dynamical one $\tau_{dyn}$ \citep{Hujeirat2000,Hujeirat2005,Orlando2005,Orlando2008}.



\subsection{Initial and boundary conditions: barotropic stellar atmosphere}
\label{sec:inicond}
Most \emph{IPCV}s have a $0.8-0.9 M_{\odot}$ white dwarf for which an appropriate radius is $R_{*}=6.0\cdot10^{8} cm$ (i.e. $10^{-2} R_{\odot}$). 
Nevertheless, being aware of this, we deliberately decided to consider a star more massive, with a mass at the limit of its class. This choice has been made in order to have a more rapidly evolving system, due to higher orbital velocity, thus reducing computational costs. After all, the mass of the central compact object has no other effect on the qualitative evolution of the system.
For these reasons, we located a compact object of mass $M_{*}\simeq1.4 M_{\odot}$ and spin period of $5.4$ minutes \citep{Patterson1994,Norton1999,Putney1999} at the origin of a 3D spherical coordinate system.

\begin{figure}
	\centering
	\subfigure[][]{\includegraphics[width=0.75\columnwidth]{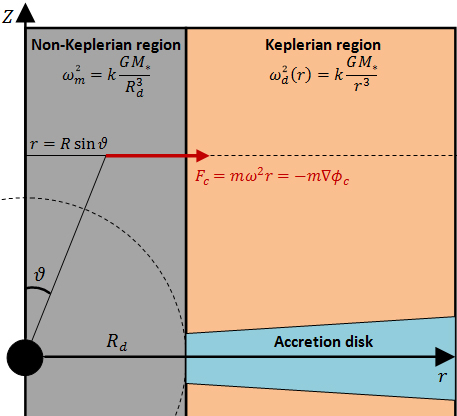}}
	\quad
	\subfigure[][]{\includegraphics[width=\columnwidth]{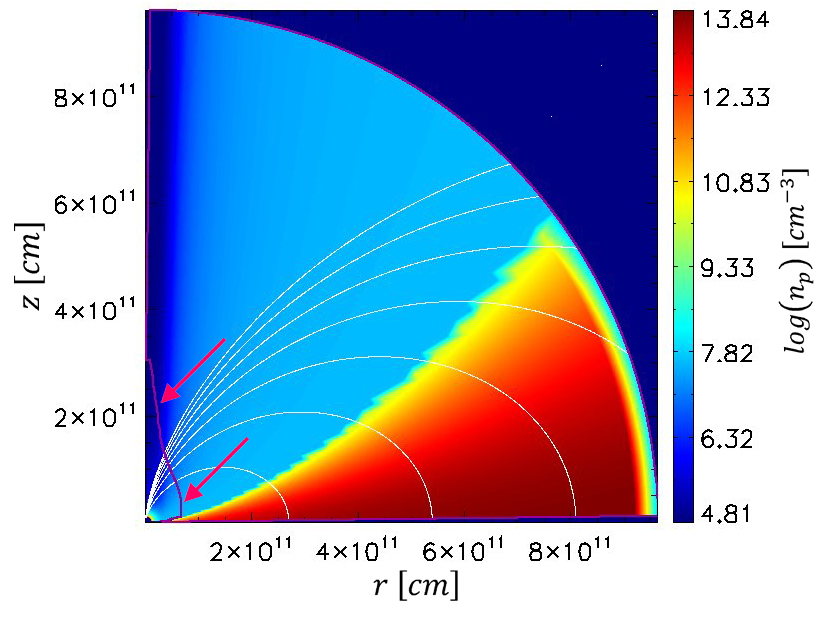}}	
	\caption{(a) Scheme of the cross section of the differentially rotating domain regions and definition of some physical quantities; (b) colour-coded log scale of density distribution in the computational domain used in the preliminary 2.5D simulation. White lines represent projected magnetic field lines, whilst the magenta line (marked by arrows) is the $\beta=1$ surface (run F4f1.00a).}
	\label{fig:2Ddomain_1}
\end{figure}

\begin{figure}
	\centering
	\subfigure[][]{\includegraphics[width=0.95\columnwidth]{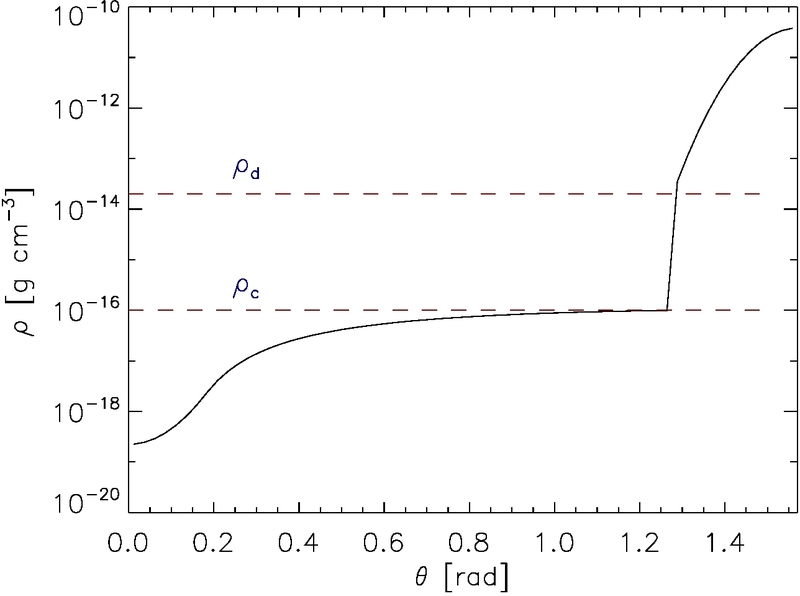}}
	\quad
	\subfigure[][]{\includegraphics[width=0.95\columnwidth]{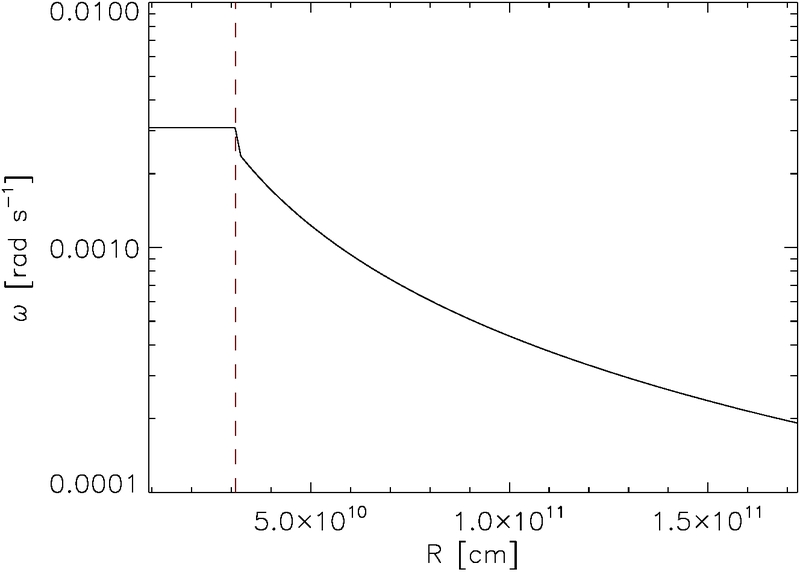}}
	\caption{Distribution of some physical quantities: (a) density profile along a spherical surface for fixed radius ($R=1.7\cdot10^{11}\;cm$) and for $\theta$ between $0^{\circ}$ (rotation axis) and $90^{\circ}$ (equatorial plane), the horizontal dashed lines show the density of the disk ($\rho_{d}$) and the corona ($\rho_{c}$) at the contact layer; (b) angular 
	velocity of the disk in the equatorial plane, the vertical dashed line identifies the truncation radius ($R_{d}$) (run F4f1.00a).}
	\label{fig:2Ddomain_2}
\end{figure}

The magnetic field of the white dwarf is supposed to be in a force-free dipole-like initial configuration \citep{Hellier2007} with the magnetic dipole aligned with the rotation axis of the star. The dipolar magnetic field may be decomposed in two components in spherical coordinate as follows:

\begin{equation}
	\begin{split}
	B_{r}      = \frac{2\mu_{B}}{r^{3}}cos\theta, 
	\qquad\qquad\qquad 
	B_{\theta} = \frac{\mu_{B}}{r^{3}}sin\theta.
	\label{eq:MagDipole}
	\end{split}
\end{equation}
\\
The magnetic moment $\mu_{B}$ has been chosen so as to get the desired magnetic field intensity at the stellar surface $B_{sup}$, ranging between $1.5\cdot10^{5} G$ and $10^{7} G$ in the several configurations explored (see Table \ref{tab:main_pars_sims}). The magnetosphere is defined as a quasi-cylindrical region of radius equal to the truncation radius $R_{d}$ where the disk structure begins in the equatorial plane. This region, surrounding the central star, is assumed to rotate as a rigid body with the same angular velocity of the star ($\omega_{*}=\omega_{mag}$) (see Fig. \ref{fig:2Ddomain_1}a and \ref{fig:2Ddomain_2}b). 

The accretion disk in our model consists of relatively cold ($T_{d}\approx5\cdot10^{4}K$) isothermal plasma rotating with angular velocity close to the Keplerian value  \citep{Orlando2011}. Disk density is structured in order to follow the barotropic prescription (see below) and has been chosen in order to get an inner disk particle density $n_{d}$ ranging between $\approx10^{10}cm^{-3}$ and $\approx10^{14}cm^{-3}$ in the various system configurations considered (see Table \ref{tab:main_pars_sims}).
The disk is initially truncated by the stellar magnetosphere at the truncation radius $R_{d}$, where there is equilibrium between plasma and magnetic pressure and therefore the ratio of the plasma pressure to the magnetic pressure, $\beta=8\pi (p + \rho v^2)/B^2$, is close to unity. In accordance with the others parameters chosen, $R_{d}$ has been placed at 
$2.7\cdot 10^{10}cm$ 
from the center of the star \citep{Hellier1993}. The corotation radius, where plasma rotates at the same angular velocity of the star, is imposed to be identical to the truncation radius, i.e. $R_{cor}=R_{d}$.
This assumption is based on the belief that most intermediate polars are in a sort of rotational equilibrium \citep{Warner1996,Norton1999}. Basically, the accretion disc is disrupted at the radius ($R_{d}$) where the Keplerian rotation of the disk is equal to that of the white dwarf, that is $\omega_{*}=\omega_{d}(R_{d})$.

The extended corona is a shell of optically thin and hot gas above the accretion disc \citep{Giovannelli2012}. In our model this region has a density following barotropic prescription (see below), whilst temperature is imposed to be initially uniform and equal to $T_{cor}=10^{7}K$, defining an isothermal low density region extending above the surface of the disk and connecting it with the central star.

In order to define a circumstellar medium in a quiescent configuration, we adopted the barotropic conditions introduced by \cite{Romanova2002}. These initial conditions satisfy mechanical equilibrium involving centrifugal, gravitational, and pressure gradient forces.
Generalizing hydrostatic balance equation, we can consider different contributions to the force, beyond the gravitational one:

\begin{equation}
	\begin{split}
	\nabla p = -\rho a_{ gen } = \rho\nabla\mathcal{F} =
	\rho\nabla\left(\phi_{Kep}-\phi_{g}-\phi_{c}\right).
	\label{eq:GenBalanceEq}
	\end{split}
\end{equation}
\\
The various contributions to the $\mathcal{F}$ function are gravitational ($\phi_{g}$), centrifugal ($\phi_{c}$), and a non-Keplerian correction ($\phi_{Kep}$). They are defined as follows:

\begin{subequations}
	\begin{align}
		&\phi_{Kep}=\left(k-1\right) \frac{GM_{*}}{R_{d}};\\[10pt] 
		&\phi_{g}  =\frac{GM_{*}}{r};\\[10pt]
		&\phi_{c}  =
		\begin{cases} 
			-\int_{\infty}^{R_{d}}\omega_{d}^{2}r\ dr &-\int_{R_{d}}^{r}\omega_{m}^{2}r\ dr \\&= k\frac{GM_{*}}{R_{d}}\left[ 1 + \frac{R_{d}^2-r^{2}}{2R_{d}^2} \right] \qquad (r\leq R_{d}) \\ \\
			-\int_{\infty}^{r}\omega_{d}^{2}r\ dr &= k\frac{GM_{*}}{r} \qquad\qquad\qquad\;\; (r\geq R_{d}).
		\end{cases}
		\label{eq:Potentials}
	\end{align}
\end{subequations}
\\
Where $r=Rsin\theta$ is the cylindrical radius and $k=1.01$ is a parameter taking into account that the disk is slightly non-Keplerian. Moreover, (see Fig. \ref{fig:2Ddomain_1}a and \ref{fig:2Ddomain_2}b) the angular velocity of the plasma is given by

\begin{equation}
	\omega^{2} =
	\begin{cases} 
	\omega_{m}^{2}=k\frac{GM_{*}}{R_{d}^{3}} \qquad (r\leq R_{d}) \\ \\
	\omega_{d}^{2}=k\frac{GM_{*}}{r^{3}} \qquad\  (r\geq R_{d}). \\
	\end{cases}
	\label{eq:AngVel}
\end{equation}
\\
In Eq. (\ref{eq:AngVel}) we distinguish $\omega_{m}$, the angular velocity of magnetosphere, where plasma is rotating as a rigid body, from $\omega_{d}$, the angular velocity of the quasi-Keplerian region. 
Using the perfect gas law $\left(p=\frac{2\rho k_{b}T}{\mu m_{H}}\right)$, we can compute the pressure at the contact layer between the corona and disk:

\begin{equation}
	\begin{split}
	p_{0} &=\frac{2\rho_{c} k_{b}T_{c}}{\mu m_{H}}\\ \\
	p_{0} &=\frac{2\rho_{d} k_{b}T_{d}}{\mu m_{H}} 
	\qquad \Rightarrow \qquad
	T_{d}=\frac{p_{0}\mu m_{H}}{2\rho_{d}k_{b}}.
	\label{eq:p0_Td}
	\end{split}
\end{equation}
\\ 
Where we take as input parameters coronal temperature $(T_{c})$, coronal density ($\rho_{c}$) and disk density ($\rho_{d}$), for which the ratio $\rho_{d}/\rho_{c}=200$ at disk surface (see Table \ref{tab:main_pars_sims}).
Finally, Eq. (\ref{eq:GenBalanceEq}) yields pressure as a function of $\mathcal{F}$

\begin{equation}
	\begin{split}
	\nabla p=\rho\nabla\mathcal{F}=p\frac{\mu m_{H}}{2k_{b}T}\nabla\mathcal{F}
	\; &\Rightarrow \; 
	\int_{p_{0}}^{p}\frac{dp}{p}=\frac{\mu m_{H}}{2k_{b}T}\int_{0}^{F}d\mathcal{F}\\ \\
	\; &\Rightarrow \;
	p=p_{0}e^{\mathcal{F}\frac{\mu m_{H}}{2k_{b}T}}.
	\label{eq:PressureF}
	\end{split}
\end{equation}
\\
and finally we get pressure and density as follows

\begin{equation}
	p =
		\begin{cases} 
			p_{0}e^{ \mathcal{F} \frac{\mu m_{H}}{2k_{b}T_{c}}}
			\qquad \text{\emph{Corona}}\\ \\
			p_{0}e^{ \mathcal{F} \frac{\mu m_{H}}{2k_{b}T_{d}}}
			\qquad \text{\emph{Disk}}
		\end{cases} 
	\rho =
		\begin{cases} 
			p\frac{\mu m_{H}}{2k_{b}T_{c}} 
			\qquad \text{\emph{Corona}}\\ \\
			p\frac{\mu m_{H}}{2k_{b}T_{d}} 
			\qquad \text{\emph{Disk}}
		\end{cases} 
	\label{eq:pressure&density}
\end{equation} 
\\
From these values we obtain an analytic function describing density and pressure in the whole domain taken into account (see Fig. \ref{fig:2Ddomain_1}b and \ref{fig:2Ddomain_2}a).

It is important to note that barotropic initial conditions provide a smooth start-up for the evolution of the system and, in particular, of the magnetic field, threading the disk and the corona. In fact, the corona above the disk rotates with the same angular velocity of the disk, and there is no contact discontinuity which is expected otherwise, if the corona were not rotating \citep{Romanova2002,Romanova2003}.

Barotropic initial conditions along with a dipole-like magnetic field configuration may be considered an useful analytic description of a quasi equilibrium system. Nevertheless, due to the interaction with the disk, where $\beta>1$ and plasma dynamics dominates, the magnetic field is strongly influenced by the Keplerian rotation of the disk. Consequently, after a certain amount of time, the magnetic field is expected to be deformed and put into a coil shape due to the interaction with the disk. 
In order to get a configuration of the system as much realistic as possible, we divided the numerical simulations in two distinct steps.

\begin{figure*}
	\centering
	\includegraphics[width=\textwidth]{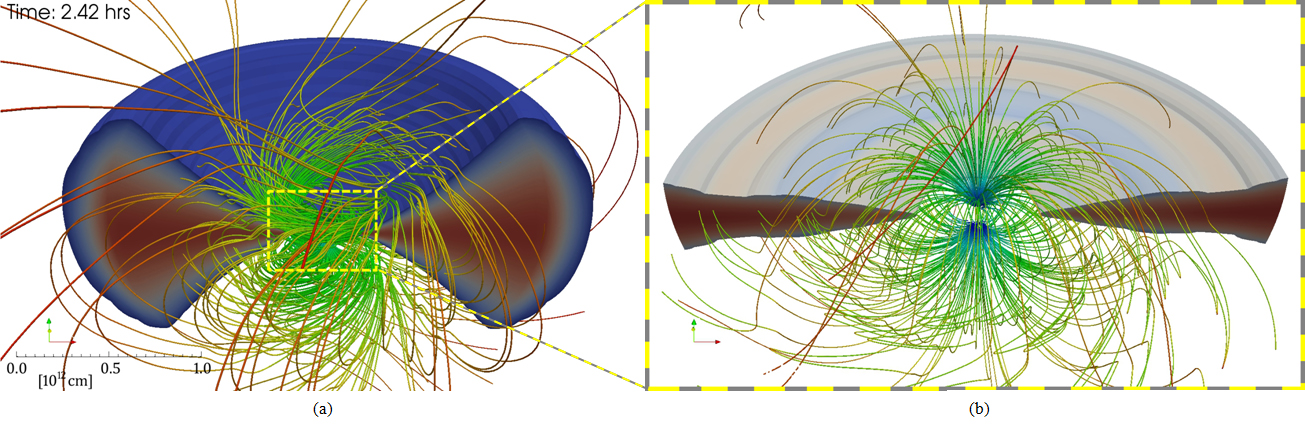}
	\caption{Result of the simulation performed to generate the initial configuration of the 3D simulation: (a) cutaway view of the 
	simulations after $\sim2.5\;hrs$ with a fully developed magnetic field, the yellow box contains the region mapped in 3D and used as initial condition of the 3D simulations; (b) rendering of the initial condition of 3D simulations, the borders of the figure are the same as those of the yellow box in the left panel (run F3f1.00c).}
	\label{fig:2Ddomain_in3D}
\end{figure*}

\begin{figure}
	\centering
	\begin{overpic}[width=0.92\columnwidth,tics=10]{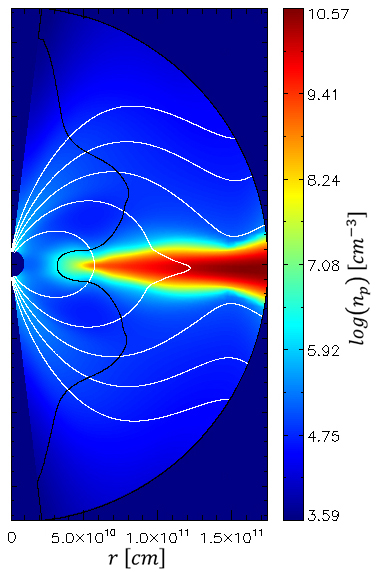}
		\put (11,81) {\textsf{\large\color{black}\textbf{$\swarrow$}}}
		\put (14,82) {\textsf{\large\color{black}\textbf{$\beta=1$}}}
		\put (27,69) {\textsf{\large\color{black}\textbf{$\beta>1$}}}
		\put (12,61) {\textsf{\large\color{black}\textbf{$\beta<1$}}}
	\end{overpic}
	\caption{Colour-coded log scale of density distribution in a slice of the computational domain used as initial condition in the 3D simulation (see Fig. \ref{fig:2Ddomain_in3D}b). White lines represent the magnetic field lines projected in the $[r,\theta]$ plane, whilst the black line is the $\beta=1$ surface with the $\beta<1$ region on the left (run F3f1.00c).}
	\label{fig:2Ddomain_beta1}
\end{figure}
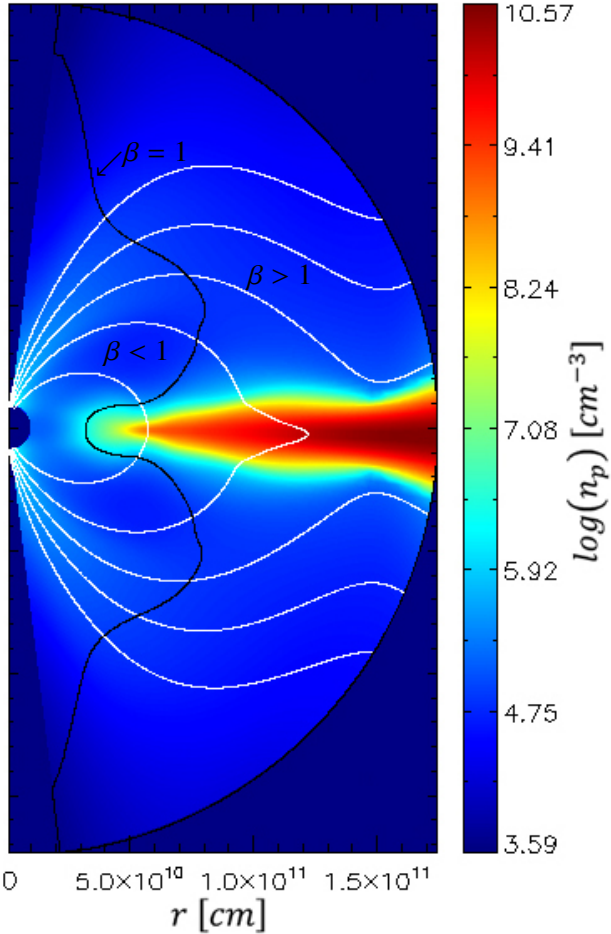

As a first step, we have run several 2.5D simulations with a square computational domain of $114\times64$ cells in spherical coordinates $\left(r,\theta\right)$ and initial conditions defined by Eq.  (\ref{eq:GenBalanceEq} - \ref{eq:pressure&density}).
The radial direction has been discretized on a grid with space steps increasing exponentially with $R$. The domain extends from 
$R_{min}\approx 9\cdot10^{9} cm$ 
to 
$R_{max}\approx 1.8\cdot10^{12} cm$. 
With this choice, the inner boundary does not coincide with the stellar surface and, therefore, the region closer to the star is not taken into account in our simulations. We made this choice for two main reasons: first, this region has not a relevant impact on our study, which is rather focused on the evolution of the region above disk surface; second, in this way we avoid computational cells having very small $\delta\theta$ mesh size, because they dramatically increase the computational time. The radial grid is made of $n_{r}=114$ points with a maximum resolution of $\Delta R=4.2\cdot10^{8} cm$, close to the star, and $\Delta R=8.2\cdot10^{10} cm$, at the external end of the domain. The disk has been externally truncated at $R=9\cdot10^{11} cm$, i.e. approximately at half radial domain. The boundary conditions at $R_{min}$ (i.e. at the end close to the star) are assumed so that infalling material follows field lines, passing through the surface of the boundary as an outflow \citep{Romanova2002}. At $R_{max}$ zero-gradient boundary condition is assumed.
As for the angular coordinate $\theta$, it ranges from $0^{\circ}$ (rotation axis) to $90^{\circ}$ (equatorial plane) and has been discretized uniformly using $n_{\theta}=64$ points, giving a resolution of $\Delta\theta\sim1.4^{\circ}$. 
For $\theta=0^{\circ}$ (i.e rotation axis) axisymmetric boundary conditions are assumed; i.e. we prescribe a sign change across the boundary of all vectorial components, except for components along the axis itself. 
For $\theta=90^{\circ}$ (i.e equatorial plane) equatorial-symmetric boundary conditions are assumed; i.e. we prescribe a sign change across the boundary of the normal component of velocity and of the component of the magnetic field tangential to the interface. 
We let the magnetic field interact with the disk for almost $2.5\;hrs$ of physical time so as to reach an almost steady state configuration.
It is worth noting that, in this preliminary step of the implemented two-phases technique, the whole size of the domain and disk has been intentionally overestimated in order to avoid undesired boundary effects during the evolution of the system. In fact, several tests showed that the prescription of a zero-gradient boundary condition (i.e. an "outflow b. c.") at the outer end of the disk might not reproduce correctly the dynamics of the plasma resulting in an underestimate of the plasma inflow. To this end, in the following phase we take into account only the inner part of the 2.5D domain (see Fig. \ref{fig:2Ddomain_in3D}).

As a second phase, we then use such a virtually relaxed and steady state as the initial conditions for our project. To this end we map the output of the 2.5D MHD simulations (namely the fully developed physical quantities: mass density, temperature, velocity and magnetic field), performed in the first phase, into a 3D domain in order to provide the initial condition of a 3D model describing the further evolution of the star-disk system (see Fig. \ref{fig:2Ddomain_in3D}).
Due to the fact that the 2.5D domain covers only a slice of half hemisphere of the system, the 2.5D-into-3D mapping has been performed through a preliminary reflection across the equatorial plane followed by a rigid rotation about the $z$ axis.
Note that the 2.5D domain has been selected in order to describe in 3D only the inner part of it, where flaring activity is expected to occur, thus reducing the computational cost. Moreover, no phenomena relevant for the present study involve the external region of the disk. 
In the initial configuration of the 3D simulations (see Fig. \ref{fig:2Ddomain_in3D}b and \ref{fig:2Ddomain_beta1}) the magnetic field lines appear still in a dipole-like form. This is particularly evident in the corona closer to the star (i.e. the magnetosphere), where $\beta<1$ and the region is field-dominated. On the other hand, in the external corona and, especially, inside the disk, where $\beta>1$, the dynamics is matter-dominated and the differential rotation of the plasma induces a noticeable deformation of the field lines.
The 3D computational domain obtained in this way is defined in spherical coordinates $\left(r,\theta,\phi\right)$ and consists of a computational cube of $128^{3}$ cells. 
The radial direction has been discretized on a grid with interval size exponentially increasing with $R$. The domain extends from 
$R_{min}\approx 9\cdot10^{9} cm$ 
to 
$R_{max}\approx 1.8\cdot10^{11} cm$. 
The radial grid is made of $n_{r}=128$ points with a maximum resolution of $\Delta R=2.1\cdot10^{8} cm$, close to the star, and $\Delta R=4.2\cdot10^{9} cm$, close to the external part of the disk. The boundary conditions at $R_{min}$ and $R_{max}$ are the same as those used in the 2.5D runs.
As for the angular coordinate $\theta$, the computational domain spans an angle from $\sim 7^{\circ}$ to $\sim 173^{\circ}$, leaving two cones above star poles as not modelled. 
Hence, the boundaries in $\theta$ do not coincide with the rotational axis of the star-disk system in order to avoid extremely small $\delta\phi$ values cause strongly increasing computational cost. 
On the other hand, the modeling of the whole computational domain, including the two cones above the star poles, has been performed in the studies of the magnetospheric accretion onto the surface of the star \citep{Romanova2003,Romanova2004,Kulkarni&Romanova2013}.
Moreover, all events relevant for the present study do not involve any region close to the rotation axis. In fact, in our problem we concentrate on the disk-magnetosphere interaction and heating of the inner disk.
The angular coordinate $\theta$ has been discretized uniformly using $n_{\theta}=128$ points, giving a resolution of $\Delta\theta\sim1.3^{\circ}$. Zero-gradient boundary conditions are assumed for $\theta$ at $\sim 7^{\circ}$ and $\sim 173^{\circ}$. Finally, in the angular coordinate $\phi$, the computational domain spans the angle from $\sim 0^{\circ}$ to $\sim 180^{\circ}$, with an uniform resolution of $\Delta\phi\sim1.4^{\circ}$. Thus the computational domain covers only half of the star-disk system. The boundary conditions for the coordinate $\phi$  are assumed to be periodic.


\subsection{Coronal heating and flaring activity}
\label{sec:flaringactivity}
The coronal heating has been prescribed as a phenomenological term, composed of two components. 
The first one works at temperature $T\leq2\;10^{4} K$ and balances exactly the local radiative losses. 
Nevertheless, several tests on the $3D$ simulations showed that increasing this value to $T\leq1\;MK$ determines a way more stable calculation and significant reduction of computational costs with comparable results in terms of dynamics of the plasma and X-ray emission. 

The second component is transient and consists of flares occurring on the surface of the disk. Every single flare is simulated and triggered injecting energy in the system through a heat pulse.
We considered the release of several flares with a randomly generated timing in order to achieve a frequency ranging from approximately 1 flare every 2.5 seconds to 1 flare every 100 seconds \citep{Tamburini2009}. 
The location of every heat pulse is randomly chosen in an angular interval $\delta\theta\approx3^{\circ}$ reckoned from the accretion disk surface. The area subject to flaring activity is the annular region between 
$5.4\cdot 10^{10} cm$ 
and 
$1.3\cdot 10^{11} cm$. 
This assumption stems from an observation concerning the measurement of the size of accretion disk coronae, based on ingress/egress timing, in another but similar context \citep{ChurchBalucinska2004}. The 3D spatial distribution of each pulse is defined to be gaussian with a width of $\sigma = 5\cdot10^{8} cm$.
The total time duration of every pulse is set at $100 s$, afterwards the pulse is completely switched off. The time evolution of pulse intensity is divided in three equally spaced parts: a linearly increasing ramp, a steady part and a linearly decreasing ramp.
The energy is injected with a maximum intensity of $H_{0} = 2\cdot10^{5} erg \cdot cm^{-3} \cdot s^{-1}$. This value of power per unit volume yields a maximum total energy of a single flare of $E_{f max} \approx 10^{34} erg$. As a conservative choice we have taken values typical of large solar flares \citep{Priest2014book}. Nevertheless the effective intensity of flares has been randomly generated with a power law with $\alpha=-2.1$ \citep{Priest2014book}, approximating the intensity distribution of flares on the Sun.


\subsection{Synthesis of the thermal X-ray emission}
\label{sec:XraySynth}
From the model results we synthesize the thermal X-ray emission originating from the disk-corona system in different spectral bands of interest: $[0.1-2.0]\;keV$ and $[2.0-10]\;keV$. We apply a method analogous to the one described by \cite{Orlando2009}.
The results of numerical simulations are the evolution of the various physical quantities of interest (density, temperature, velocity and magnetic field) of the plasma in half of the spatial domain.
In order to get the total emitted radiation, we reconstruct the 3D spatial distribution of these physical quantities in the whole spatial domain. We rotate the system around the $x$ axis to explore different inclinations of the orbital plane relative to the sky plane. 
From the value of temperature and emission measure of the \mbox{$i$-th} computational cell we synthesize the corresponding thermal X-ray emission, using the Astrophysical Plasma Emission Code (APEC) \citep{Smith2001} of hot collisionally ionized plasma. 
APEC models provide emissivity tables of both line and continuum emission which may be used to calculate predicted fluxes. These predicted fluxes may, in turn, be compared with observed spectral features. The predicted flux is given by:

\begin{equation}
	F=\frac{\epsilon(T)}{4\pi R^{2}}\int n_{e}n_{H}dV \quad\Rightarrow\quad F_{i}=\frac{\epsilon(T_{i})}{4\pi R^{2}_{i}}n^{2}_{i}V_{i},
\label{eq:APECemiss1}
\end{equation} 
\\
where $\epsilon(T)$ is the emissivity in $erg\;cm^{3}/s$, $R$ is the distance to the source in $cm$, and the integral is the emission measure in $cm^{-3}$. Therefore, defining $i$ as the index of a cell of the computational domain, $\epsilon(T_{i})$ is the emissivity of the cell at temperature $T_{i}$, whilst the emission measure is defined as $EM_{i}=n_{i}^2 V_{i}$, where $n_{i}$ is the proton number density in the cell and $V_{i}$ is the volume of the cell where plasma is assumed to be fully ionized. The spectral synthesis takes into account the Doppler shift of emission lines due to the plasma velocity component along the line of sight (LoS).
We assume solar metal abundances of \cite{AndersGrevesse1989} for the circumstellar medium (CSM).
The X-ray spectrum emitted by each computational cell is filtered through the CSM absorption column, going from the cell to the observer along the LoS. The CSM absorption is computed using the absorption cross-section as a function of wavelength from \cite{Balucinska1992}.
The thermal X-ray emission, integrated along the LoS in the whole computational domain, defines an image describing the distribution of the thermal X-ray flux; the flux values are computed assuming a distance of 500 pc and no interstellar absorption. Then, through the integration of the fluxes of the images, we obtain the total emission at the selected time; the sequence of all these values at different times yields the lightcurve in the selected energy bands.

\section{Results}
\label{sec:Results}


The model described in Section \ref{sec:Model} allows to explore different configurations of the system changing some main input parameters. In fact, once we set the truncation radius at $R_{d}$, where the $\beta$ of plasma is imposed equal to unity, we can change the value of density at the surface of the disk and, consequently, obtain the surface magnetic field of the white dwarf needed to keep $\beta=1$ at the truncation radius $R_{d}$. The various configurations of the system taken into account have been chosen in order to have a sampling as wide as possible of magnetic field strength found in this kind of systems, that ranges between $0.1$ and $10\;MG$. We take as a reference model the one defined as F4f1.00a in Table \ref{tab:main_pars_sims}, in which the surface magnetic field intensity is $\approx5\;MG$. Other parameters have been taken into account in our exploration of the parameter space. In particular we explored the maximum intensity of the flares and their frequency as well. Table \ref{tab:main_pars_sims} summarises the various simulations and details the main parameters characteristic of each run: 
$\rho_{c}$ and $\rho_{d}$ are, respectively, coronal density and disk density at the layer between corona and disk; 
$T_{c}$ is the initial uniform temperature of the corona;
$n_{d}$ is the maximum density of particles in the bulk of the disk; 
$B_{sup}$ is the magnetic field intensity at the surface of the star;
$FF$ is the frequency of flare release, and finally
$I_{max}$ is the maximum energy release of the flares.

\begin{table*}\footnotesize
	\caption{Main input parameters defining initial condition described in section \ref{sec:inicond}.} 
	\label{tab:main_pars_sims} 
	\centering      
	\begin{tabular}{c c c c c c c c} 
		\hline\hline\\         
		Run                                   & 
		$\rho_{c}    \;[g/cm^{3}]$ &
		$\rho_{d}    \;[g/cm^{3}]$ &
		$T_{c}       \;[K]$        &
		$n_{d}       \;[cm^{-3}]$  & 
		$B_{sup}     \;[G]$        &
		$FF          \;[s^{-1}]$   &
		$I_{max}     \;[erg\;cm^{-3}s^{-1}]$ \\\\
		\hline\\ 
		F3f1.00c & $1.0\;10^{-19}$ & $2.0\;10^{-17}$ & $10^{7}$ & $1.8\;10^{10}$ & $1.6\;10^{5}$ & $1/5$ & $10^{34}$\\
		F6f1.00a & $1.0\;10^{-17}$ & $2.0\;10^{-15}$ & $10^{7}$ & $1.8\;10^{12}$ & $1.6\;10^{6}$ & $1/5$ & $10^{34}$\\
		F4f1.00a & $1.0\;10^{-16}$ & $2.0\;10^{-14}$ & $10^{7}$ & $1.8\;10^{13}$ & $5.0\;10^{6}$ & $1/5$   & $10^{34}$\\
		F4f1.00e & $1.0\;10^{-16}$ & $2.0\;10^{-14}$ & $10^{7}$ & $1.8\;10^{13}$ & $5.0\;10^{6}$ & $1/10$  & $10^{34}$\\
		F4f0.01f & $1.0\;10^{-16}$ & $2.0\;10^{-14}$ & $10^{7}$ & $1.8\;10^{13}$ & $5.0\;10^{6}$ & $1/2.5$ & $10^{32}$\\		
		F4f1.00f & $1.0\;10^{-16}$ & $2.0\;10^{-14}$ & $10^{7}$ & $1.8\;10^{13}$ & $5.0\;10^{6}$ & $1/2.5$ & $10^{34}$\\	
		F4f1.00g & $1.0\;10^{-16}$ & $2.0\;10^{-14}$ & $10^{7}$ & $1.8\;10^{13}$ & $5.0\;10^{6}$ & $1/100$ & $10^{34}$\\	
		F4f10.0g & $1.0\;10^{-16}$ & $2.0\;10^{-14}$ & $10^{7}$ & $1.8\;10^{13}$ & $5.0\;10^{6}$ & $1/100$ & $10^{36}$ \\
		F5f1.00a & $4.0\;10^{-16}$ & $8.0\;10^{-14}$ & $10^{7}$ & $7.1\;10^{13}$ & $1.0\;10^{7}$ & $1/5$ & $10^{34}$\\\\
		\hline 
	\end{tabular}
	\tablefoot{Coronal density $\rho_{c}$ and disk density $\rho_{d}$ refer to the values at the layer between corona and disk. The density of particles $n_{d}$ is the maximum density in the bulk of the disk. FF is the frequency of flare relase.}
\end{table*}


\subsection{System dynamics}

\begin{figure*}[h!]

	\begin{overpic}
		[trim={65px 45px 180px 170px},clip,width=0.23\textwidth,tics=10]
		{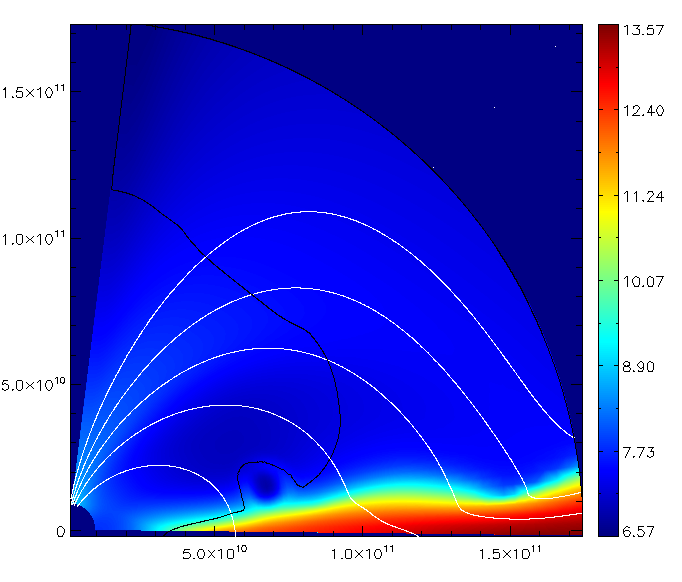}
		\put(2,77) {\textsf{\small\color{white}\textbf{Dens. \bm{$\log{n_{p}} \;[cm^{-3}]$}}}}
		\put(73,77){\textsf{\small\color{white}\textbf{\bm{$t=31\;s$}}}}
		\put(55,50){\textsf{\large\color{black}\textbf{$\swarrow$}}}
		\put(65,55){\textsf{\large\color{black}\textbf{$\beta=1$}}}
	\end{overpic}
	\begin{overpic}
		[trim={65px 45px 180px 170px},clip,width=0.23\textwidth,tics=10]
		{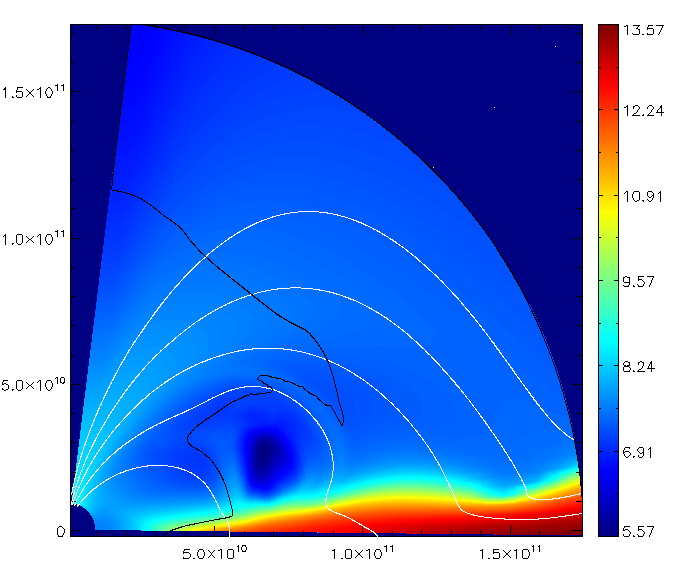}
		\put(73,77){\textsf{\small\color{white}\textbf{\bm{$t=93\;s$}}}}
	\end{overpic}	
	\begin{overpic}
		[trim={65px 45px 180px 170px},clip,width=0.23\textwidth,tics=10]
		{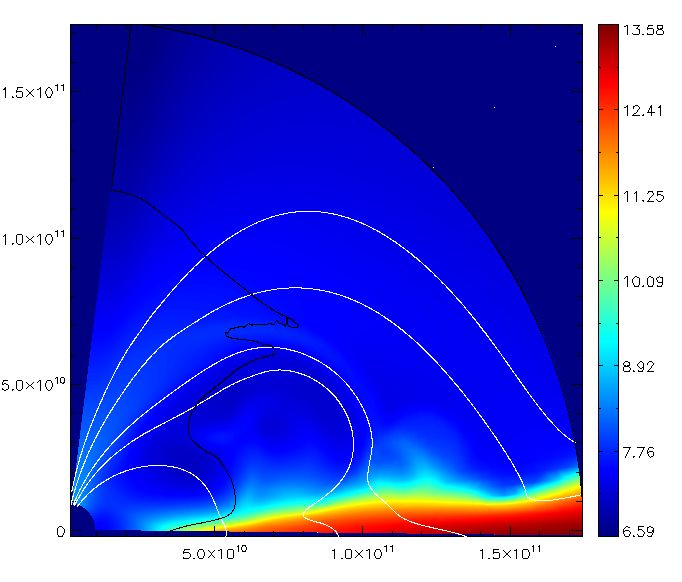}
		\put(70,77){\textsf{\small\color{white}\textbf{\bm{$t=155\;s$}}}}
	\end{overpic}	
	\begin{overpic}
		[trim={65px 45px 180px 170px},clip,width=0.23\textwidth,tics=10]
		{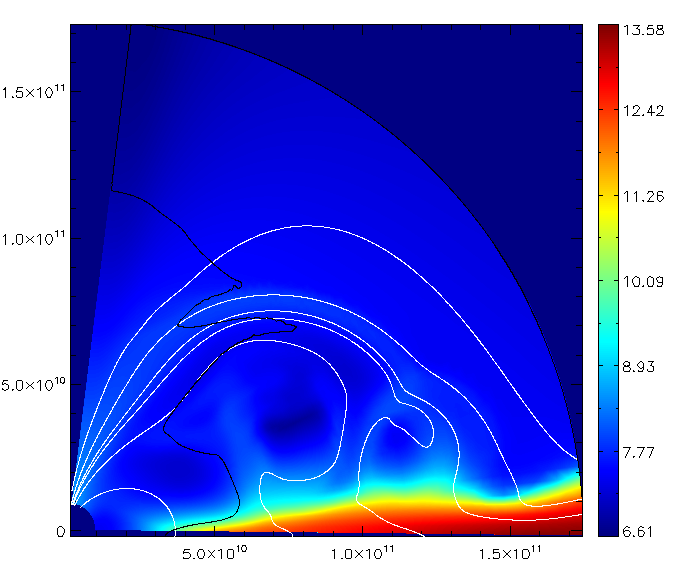}
		\put(70,77){\textsf{\small\color{white}\textbf{\bm{$t=218\;s$}}}}
	\end{overpic}
	\includegraphics[trim={600px 209px 0 0},clip,height=0.21\textwidth]
	{rho035.png}			
	
	\vspace{1mm}
	\begin{overpic}
		[trim={65px 45px 180px 170px},clip,width=0.23\textwidth,tics=10]
		{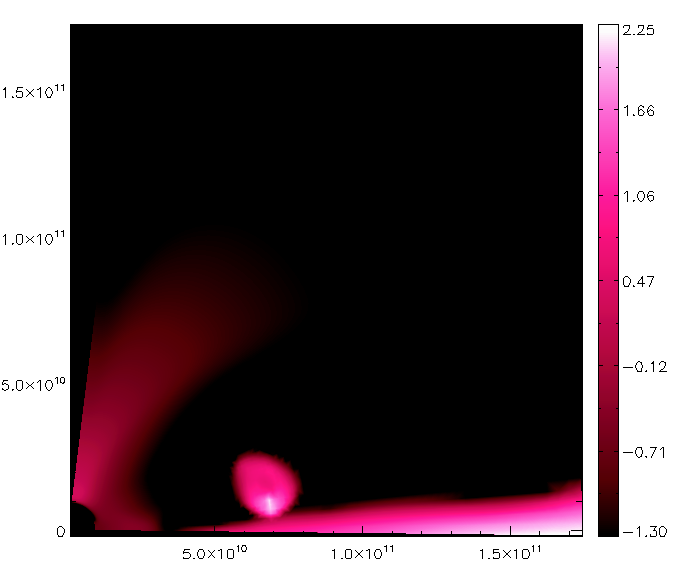}
		\put(2,77) {\textsf{\small\color{white}\textbf{Pres. \bm{$[dyne/cm^2]$}}}}
		\put(50,18){\textsf{\normalsize\color{white}\textbf{$\swarrow$}}}
		\put(58,25){\textsf{\normalsize\color{white}\textbf{1\bm{$^{st}$} flare}}}
	\end{overpic}
	\begin{overpic}	
		[trim={65px 45px 180px 170px},clip,width=0.23\textwidth,tics=10]
		{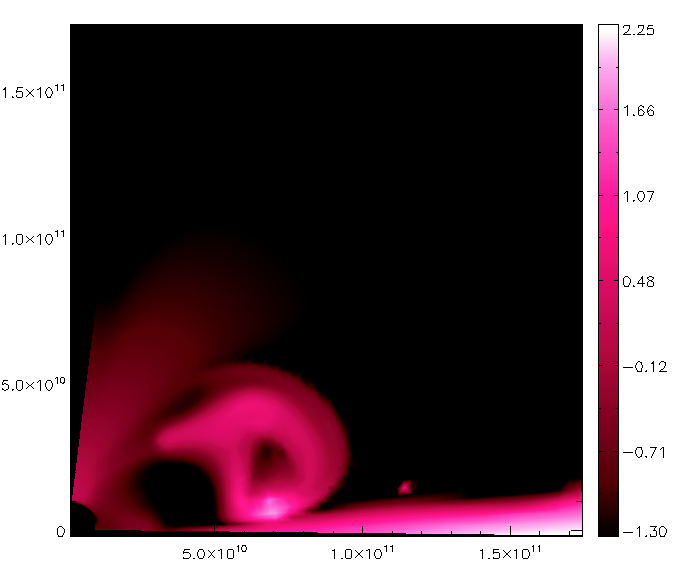}
		\put(65,23){\textsf{\normalsize\color{white}\textbf{2\bm{$^{nd}$} flare}}}
		\put(76,15){\textsf{\normalsize\color{white}\textbf{$\downarrow$}}}
	\end{overpic}	
	\includegraphics[trim={65px 45px 180px 170px},clip,width=0.23\textwidth]
	{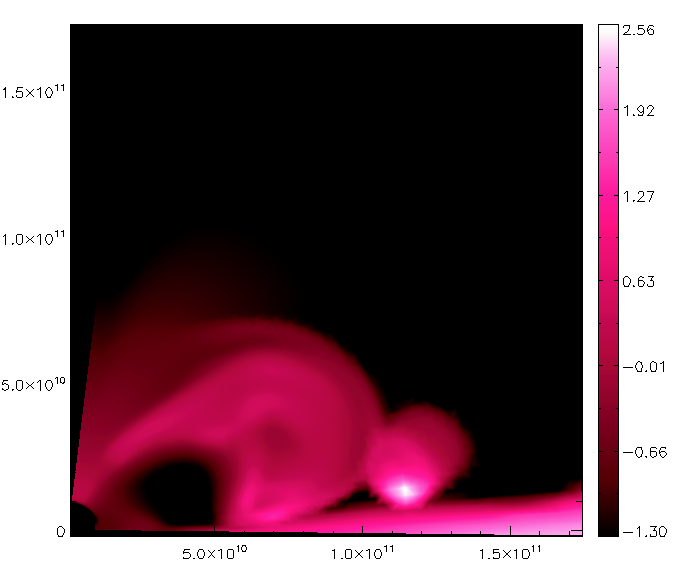}
	\includegraphics[trim={65px 45px 180px 170px},clip,width=0.23\textwidth]
	{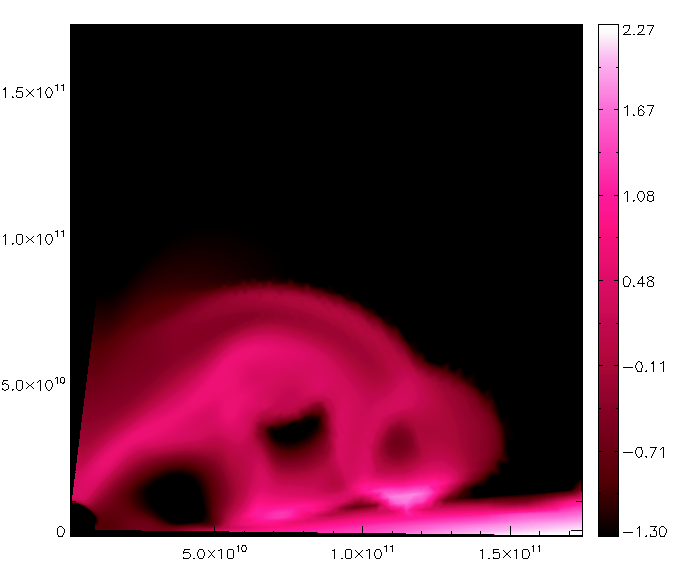}
	\includegraphics[trim={600px 209px 0 0},clip,height=0.21\textwidth]
	{prs035.png}		
	
	\vspace{1mm}
	\begin{overpic}
		[trim={65px 45px 180px 170px},clip,width=0.23\textwidth,tics=10]
		{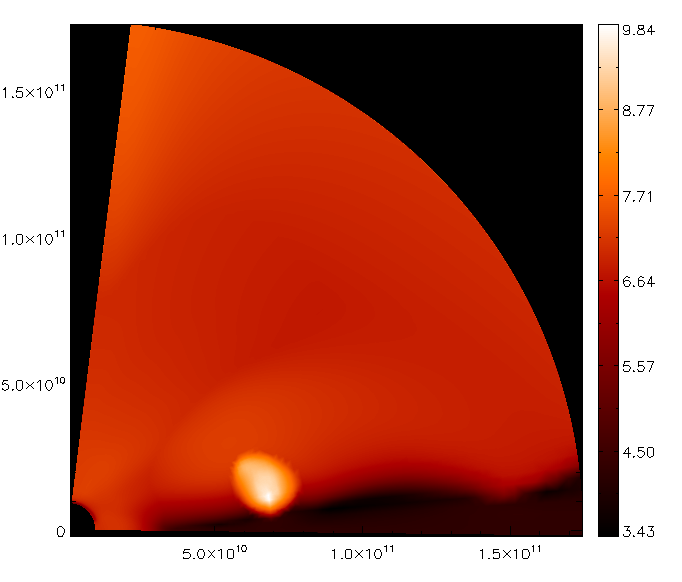}
		\put(2,77){\textsf{\small\color{white}\textbf{Temp. \bm{$\log{T}\;[K]$}}}}
		\put(0,-8){\textsf{\small\color{black}$0\;[cm]\qquad5e10\qquad1e11\qquad1.5e11$}}
	\end{overpic}	
	\includegraphics[trim={65px 45px 180px 170px},clip,width=0.23\textwidth]
	{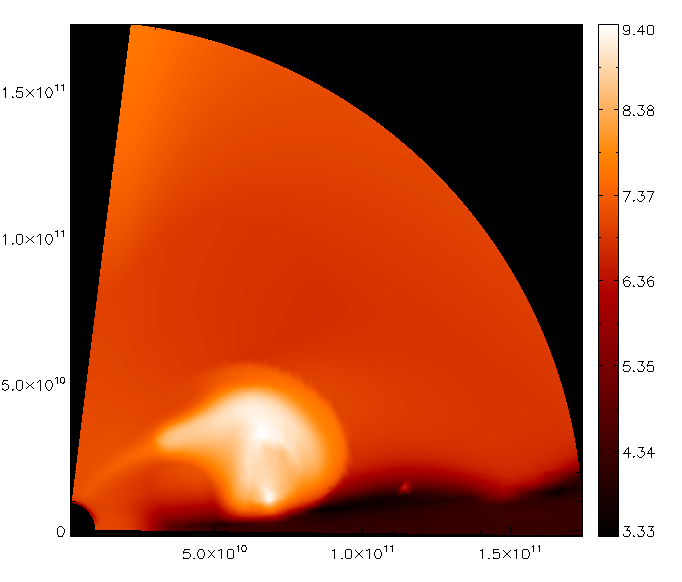}
	\includegraphics[trim={65px 45px 180px 170px},clip,width=0.23\textwidth]
	{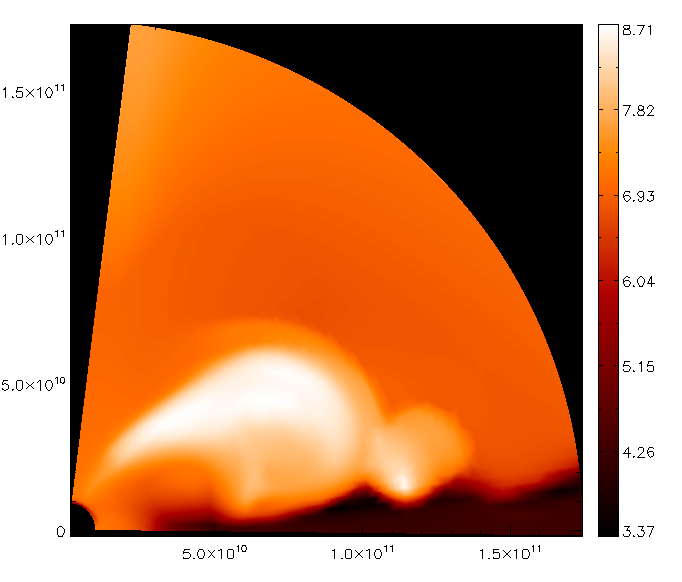}
	\includegraphics[trim={65px 45px 180px 170px},clip,width=0.23\textwidth]
	{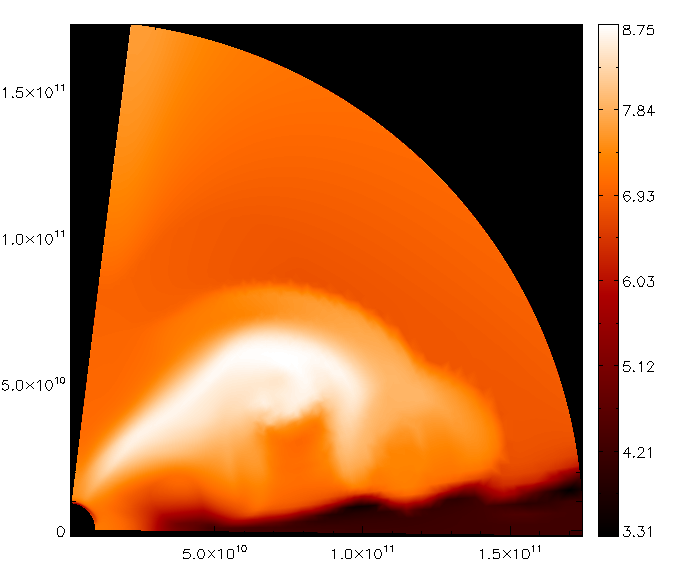}	
	\includegraphics[trim={600px 209px 0 0},clip,height=0.21\textwidth]
	{tmp035.png}				
	
	\vspace{3mm}
	\caption{Two following flaring loops during their evolution. The figure shows the distribution of density (upper panels), pressure (middle panels) and temperature (lower panels) in a slice in the $[r,\theta]$ plane passing through the middle of the loops. In the upper panels white lines represent magnetic field lines projected in the $[R,\theta]$ plane, while black line shows the $\beta=1$ surface. The slice encompasses an angular sector going from the rotation axis to the disk equatorial plane. 
	The first flare occurs in a $\beta>1$ region, surrounded with $\beta<1$ region,	and produces a magnetically confined loop linking star and disk.
	The second flare is also released in a $\beta>1$ region, but at larger distance from the star and
	at approximately the same angle $\phi$, compared with the first flare.
	It appears a bit later in time, at time $t=93s$, and becomes bright at $t=155s$. Differently from the first flare, the hot plasma of the second flare is not magnetically confined (run F4f1.00a).}
	\label{fig:Loop_slice}
\end{figure*}

For our reference case, F4f1.00a, the energy release of the intense flaring activity has been followed for approximately 20 minutes. 
Each heat pulse is injected into the system through the parametric term $Q(r,\theta,\phi,t)$ in Eq.~(\ref{eq:EneCon}) at randomly chosen locations on both the upper and lower surfaces of the disk, and triggers a MHD shock wave that develops above the disk, preferentially propagating away from the central star, in the region where $\beta>1$.
The abrupt release of energy determines a local increase of temperature and pressure, heating the dense plasma of the disk. 
While each flare develops, the heat pulse determines an overpressure in the region of the disk corresponding to the footpoint of the loop\footnote{The footpoint of a magnetic loop is the region at which tubes of magnetic field lines reach the surface of the disk to form coronal loops.}. 
This overpressure travels as a pressure wave through the bulk of the disk, eventually reaching the opposite surface of the disk.
The heated disk material expands in the magnetosphere propagating above the disk with a strong evaporation front moving at speed of the order of $10^{8}\;cm/s$. 
The heated and expanding plasma interacts with magnetic field, with mutual effects. As a matter of fact, during the evolution of the system, the field lines, in regions where $\beta$ has high values, are dragged away and slowly deformed by the action of the flares, and, eventually, only the magnetosphere close to the star keeps an approximately dipolar shape. Analogously, the hot evaporation front triggered by each flare is strongly affected by the interaction with the magnetosphere, and, where $\beta<1$, the hot plasma is channelled along the magnetic field lines toward the central white dwarf. A few minutes after each energy release, a hot $(T\sim\;10^{9}K)$ magnetic loop of length of the order of $10^{10} cm$ is formed, linking the disk with the star. 
The loop continues to expand for almost 8 minutes, and the temperature decreases by almost 2 orders of magnitude. 
The frames in Fig.~\ref{fig:Loop_slice} show the evolution of density, pressure and temperature distribution of two successive flares in a slice of the domain perpendicular to the equatorial plane and aligned with the symmetry axis. 
The first flare occurs in a low $\beta$ region closer to the star, whereas the second one occurs in a high $\beta$ region. Both flares are triggered at approximately the same angle $\phi$. For the first flare, the magnetic field, due to a low $\beta$, produces the channelling of the hot plasma, forming a magnetic loop linking the star and disk, while the the second one appears as not magnetically confined.
The plasma starts cooling as soon as the heat deposition is over, as a result of the combined action of the efficient radiative cooling and thermal conduction. 
In particular, on one hand, thermal conduction from the outer layer of hot plasma sustains the disk evaporation that goes on even after the flare decays.
On the other hand, thermal conduction, due to its high anisotropy in presence of the magnetic field (see Eq.~\ref{eq:ThermCond_q_spi}), promotes the development of a hot magnetic tube (loop) linking star and disk, through the formation of a fast thermal front propagating along the magnetic field lines toward the star and reaching it in a time-scale of $\approx1\;min$ (lower panels of Fig.~\ref{fig:Loop_slice}).

\begin{figure}
	\centering
	\includegraphics[trim={0 0 0 0},clip,width=\columnwidth]{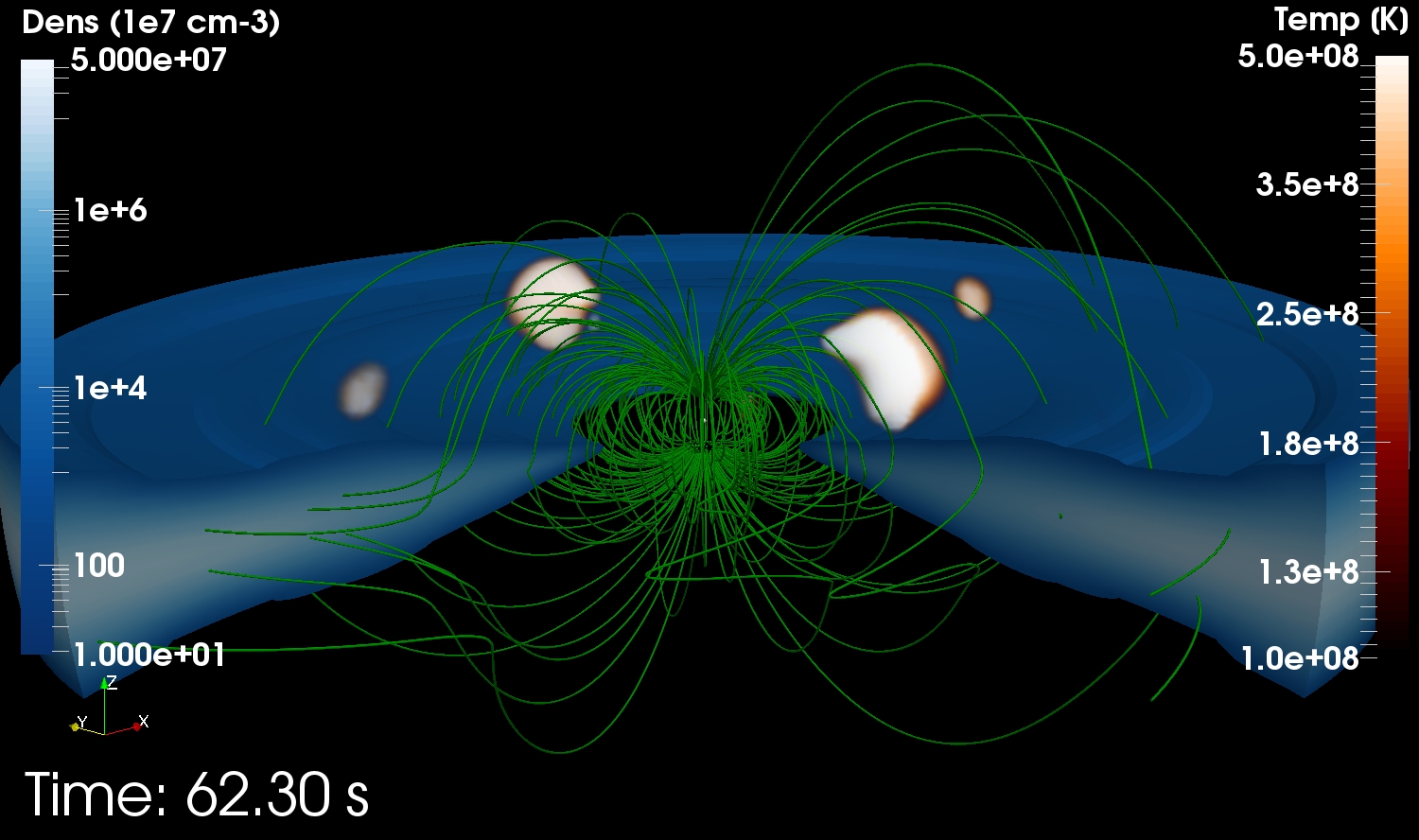}
	
	\vspace{1mm}		
	\includegraphics[trim={0 0 0 0},clip,width=\columnwidth]{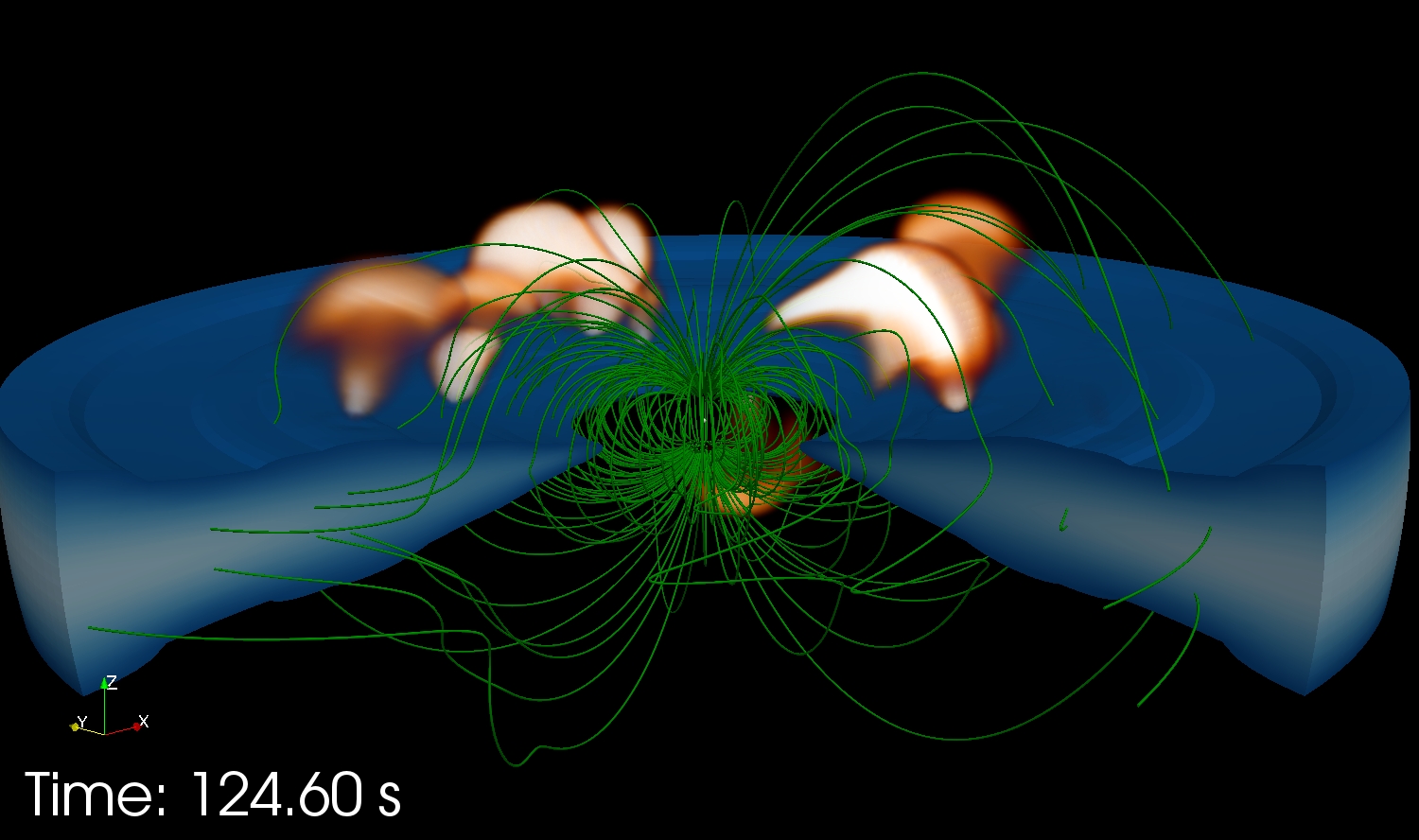}
	
	\vspace{1mm}
	\includegraphics[trim={0 0 0 0},clip,width=\columnwidth]{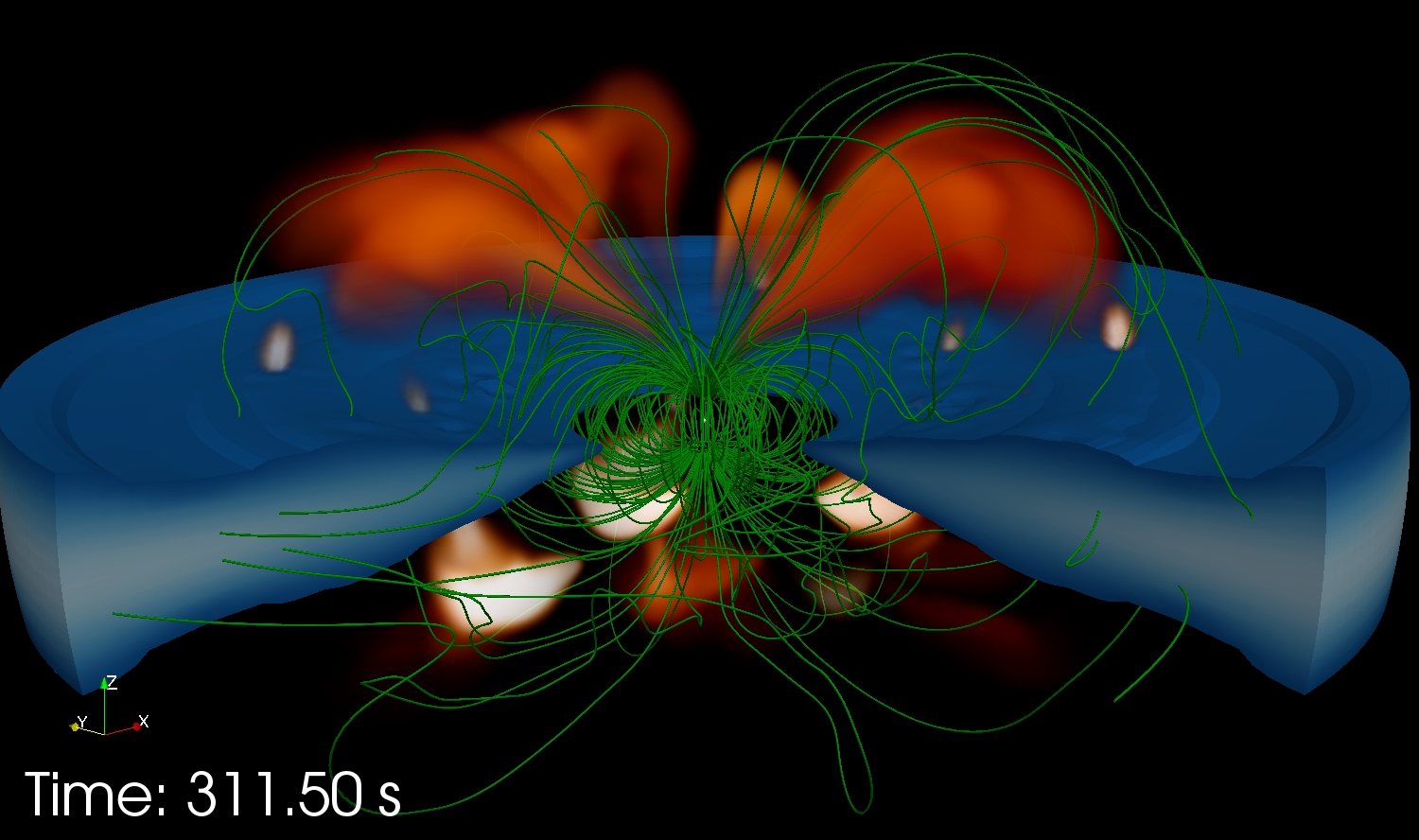}
	
	\vspace{1mm}
	\includegraphics[trim={0 0 0 0},clip,width=\columnwidth]{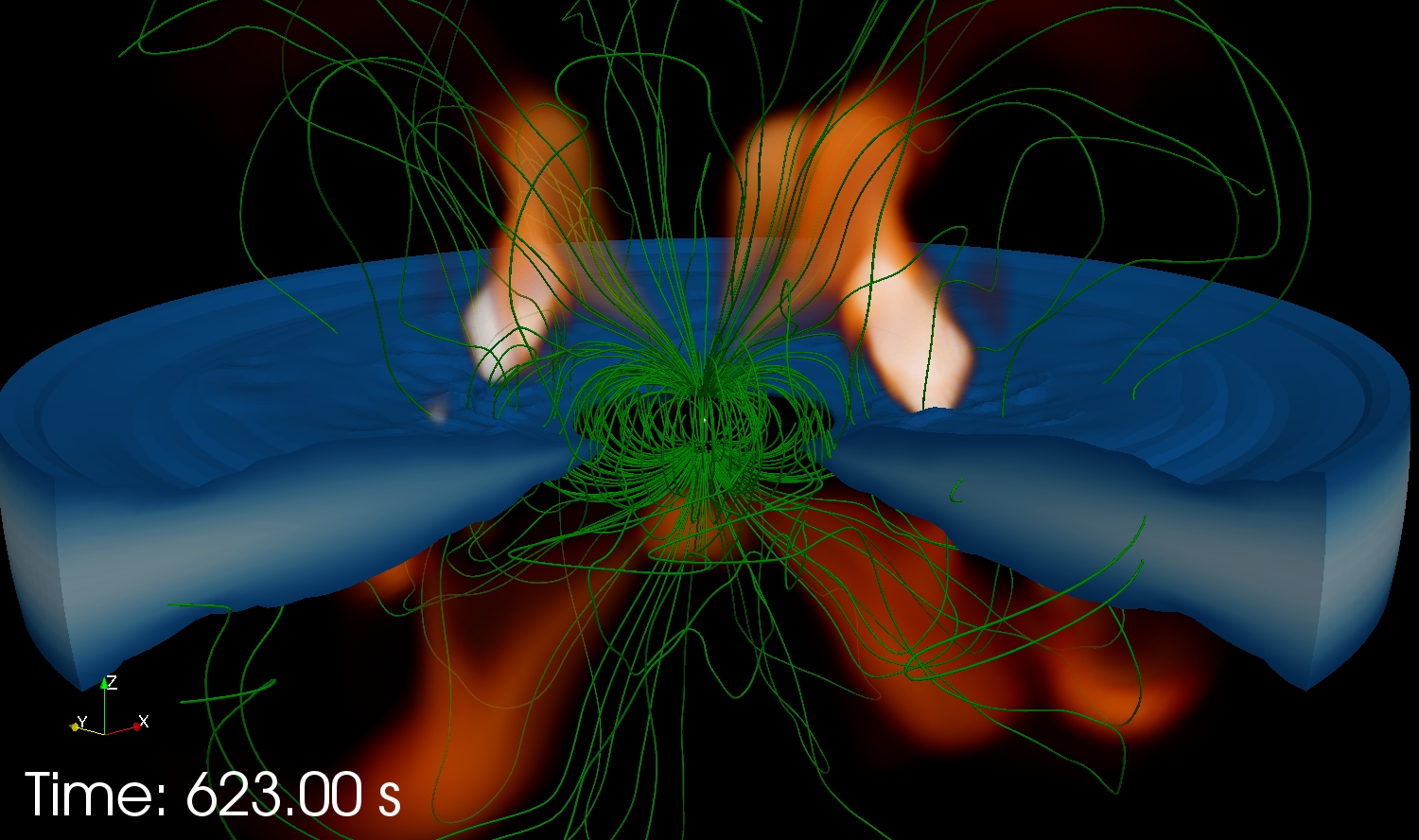}
	
	\caption{Effect of a hail of flares on the disk and on the circumstellar atmosphere.  The cutaway views of the star-disk system show the mass density (blue) and the magnetic field lines (green) at different times. The 3D volume rendering of the plasma temperature is over-plotted, showing the flaring loops (red-yellow) linking the inner part of the disk with the central white dwarf (run F4f1.00a).} 
	\label{fig:CoronalLoops}
\end{figure}

\begin{figure}
	\centering
	\begin{overpic}[width=\columnwidth,tics=10]{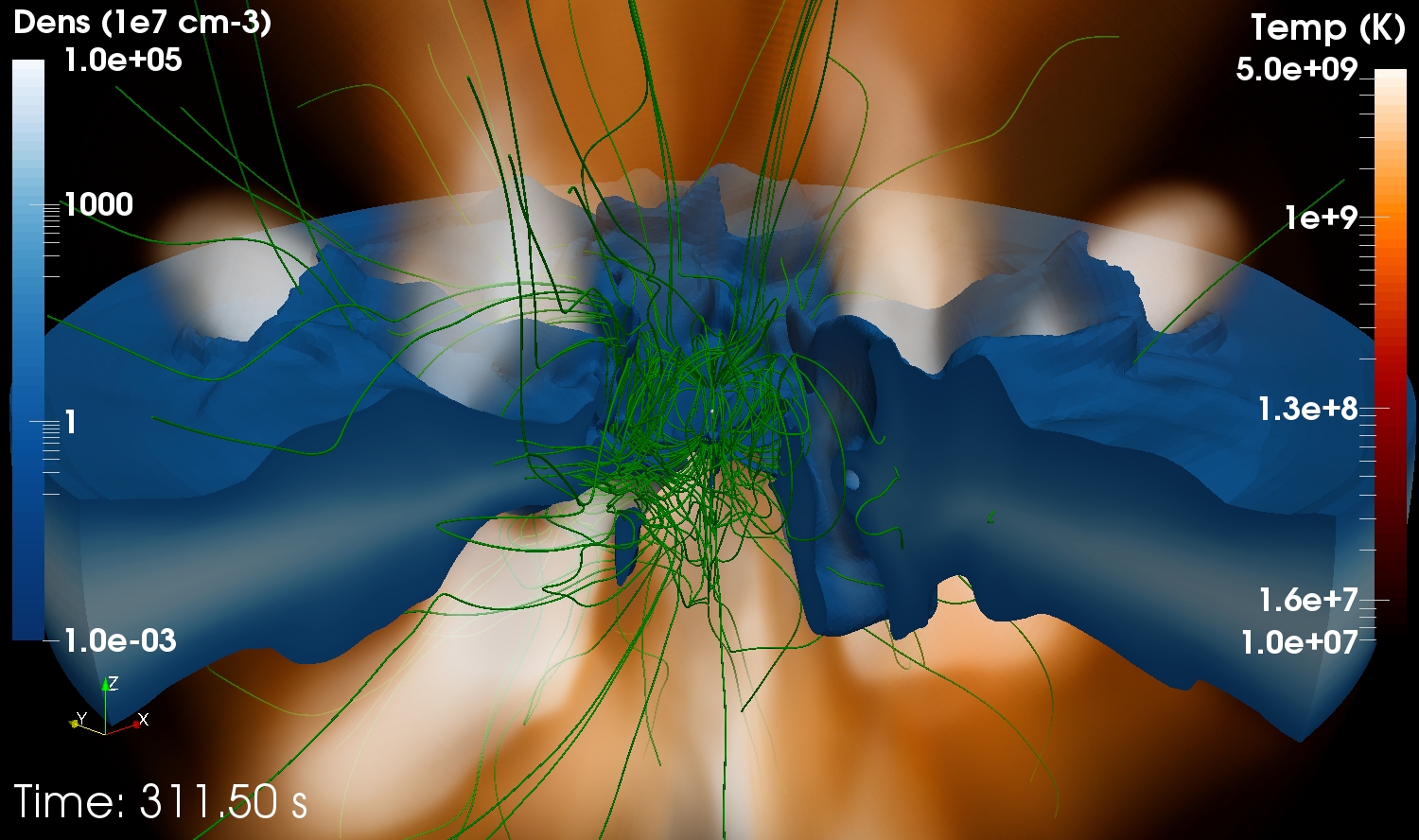}
		\put (35,55) {\textsf{\large\color{white}\textbf{Run F3f1.00c}}}
	\end{overpic}
	
	\vspace{1mm}
	\begin{overpic}[width=\columnwidth,tics=10]{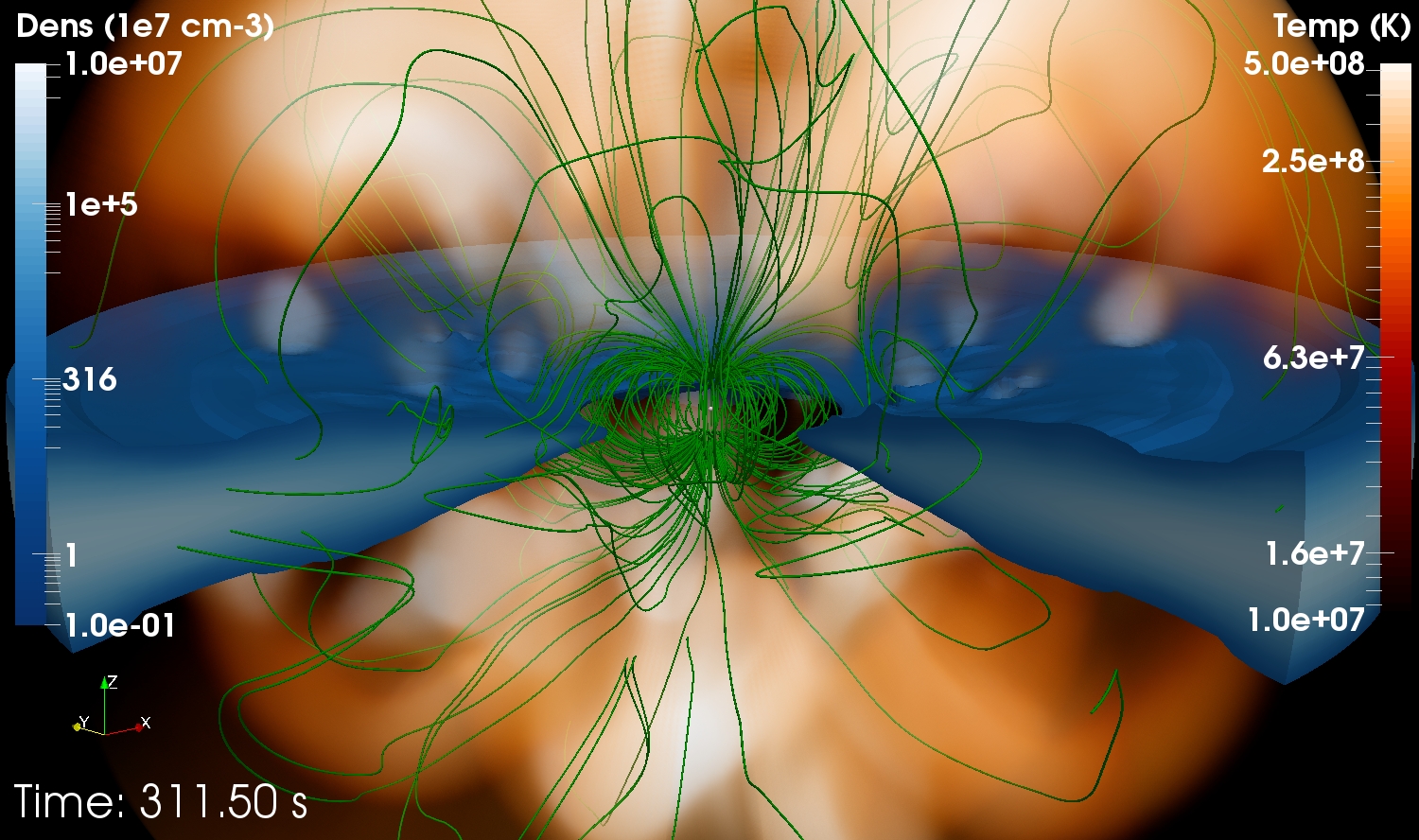}
		\put (35,55) {\textsf{\large\color{white}\textbf{Run F6f1.00a}}}
	\end{overpic}
	
	\vspace{1mm}	
	\begin{overpic}[width=\columnwidth,tics=10]{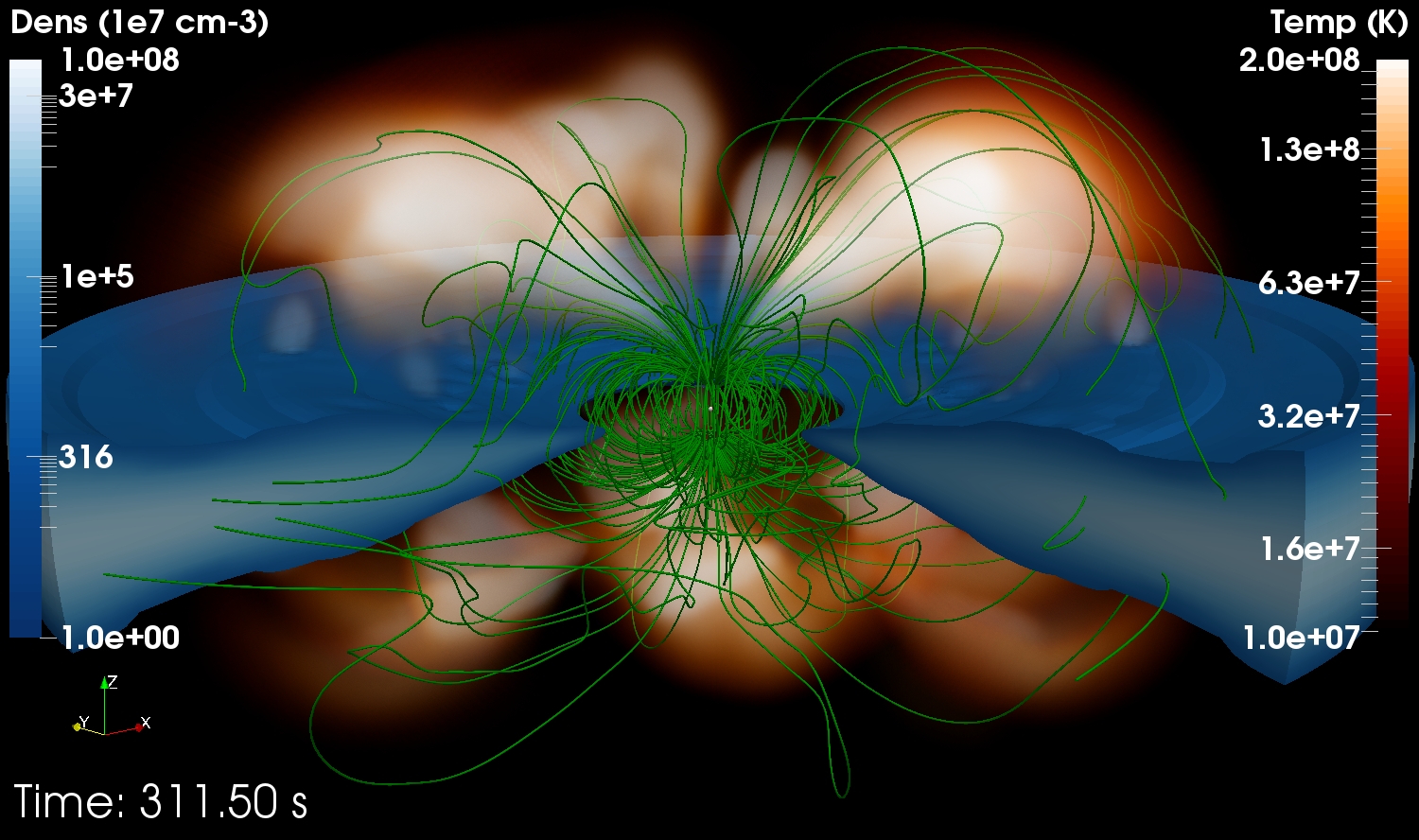}
		\put (35,55) {\textsf{\large\color{white}\textbf{Run F4f1.00a}}}
	\end{overpic}
	
	\vspace{1mm}	
	\begin{overpic}[width=\columnwidth,tics=10]{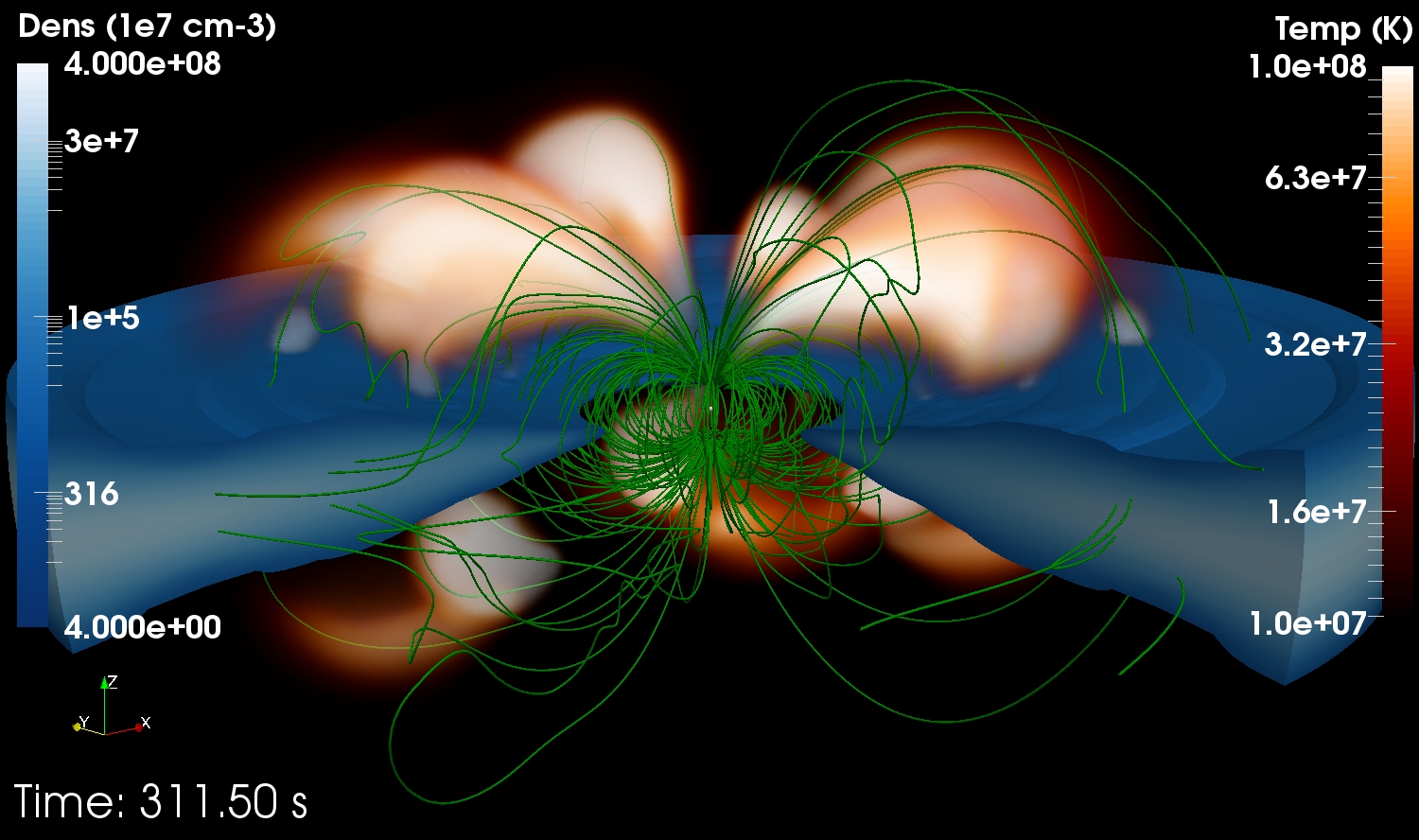}
		\put (35,55) {\textsf{\large\color{white}\textbf{Run F5f1.00a}}}
	\end{overpic}
	
	\caption{The above images show the different effects of the same energy release at the same time ($t=311.50\;s$) on different configurations of the system with progressively denser disks and stronger magnetic field, from top to bottom. The cutaway views of the star-disk system show the mass density (blue) and the magnetic field lines (green). The 3D volume rendering of the plasma temperature is over-plotted, showing the flaring loops (red-yellow).} 
	\label{fig:3Dmodels_comp}
\end{figure}

In Fig. \ref{fig:CoronalLoops} a 3D rendering of the evolution of the system is represented. 
During the system evolution, the combined effect of the storm of flares causes the formation of an extended corona between disk and star (see panels in Fig. \ref{fig:CoronalLoops}). Nevertheless, most of the evaporated disk material, particularly in the external regions of the disk, is not completely confined by magnetic field and is not efficiently channelled into the loops but it escapes upward and is ejected away in the outer regions, carrying away mass and angular momentum.

The main difference between the various models explored lies in the density of the circumstellar atmosphere and in the magnetic field intensity.
It is worth noting that $\beta$ must be always greater than unity inside the disk. This implies that, for example, in a low density disk the magnetic field intensity is correspondingly low, and the energy release during the reconnection process is also smaller. Therefore, low-density disk could not be capable of generating strong flares. With an analogous reasoning, the reverse occurs in high density configurations.
For this reason, in all the configurations explored we have always taken into account values of magnetic field intensity and disk density to keep $\beta>1$ inside the disk.
Keeping this crucial caution in mind, we explored disks with different density, however, we kept the maximum energy released by flares the same, in order to obtain comparable simulations of the various configurations taken into account.
Due mainly to a denser disk, the effects of the pressure wave moving across the disk caused by each flare are mild in configurations with denser disk and stronger magnetic field (i.e. runs F4f1.00a and F5f1.00a, see bottom panels in Fig. \ref{fig:3Dmodels_comp}). In these two cases, the perturbation of the disk is limited to the surface of the disk and the field lines are slowly deformed by the action of the MHD shock.
On the other hand, the flares cause a perturbation that is more evident in low density/weak field configurations (i.e. runs F3f1.00c and F6f1.00a, see top panels in Fig. \ref{fig:3Dmodels_comp}). Moreover, in these configurations, the perturbation of the disk causes a fast and evident deformation of the field lines. 
So, due to a less dense disk, the flaring activity can strongly alter the configuration of the disk potentially hampering its overall stability. 

Another important element differentiating the various models is how often and how intense is the energy release associated to each flare. As expected, frequency and intensity of the flares have remarkable effects on the disk and on the magnetic field lines. In fact, a higher flare frequency or a more intense energy release produces a noticeable disk disruption along with a deeper dislocation of the magnetic field lines, compared to analogous models characterized by a less intense and less frequent flaring activity. 

It is worth noting that we have simulated an intense flaring activity through randomly parametrized heat releases, thus flares do not develop as a result of the dynamics of the system, although their presence is justified by the magnetic interaction in a turbulent and differentially rotating disk. 
In our model, the seed magnetic field in the disk is due to the presence of a magnetized WD. 
However, as mentioned in Sec.~\ref{sec:Intro}, a magnetized central object is not required to the formation of an extended disk corona. 
In presence of a non-magnetized central object, we expect that loops may form as a consequence of the phenomenon of the expulsion of magnetic field from the disk \citep{GaleevRosnerVaiana1979}, therefore linking different regions of the disk. 
We conclude, therefore, that the only significant difference of the cases explored here with those considering a non-magnetized central object is the presence of magnetic loops linking the disk to the star, as observed in our simulations.


\subsection{Thermal X-ray emission}
\begin{figure}[t!]
	\centering
	\includegraphics[trim={0 0 0 0},clip,width=\columnwidth]{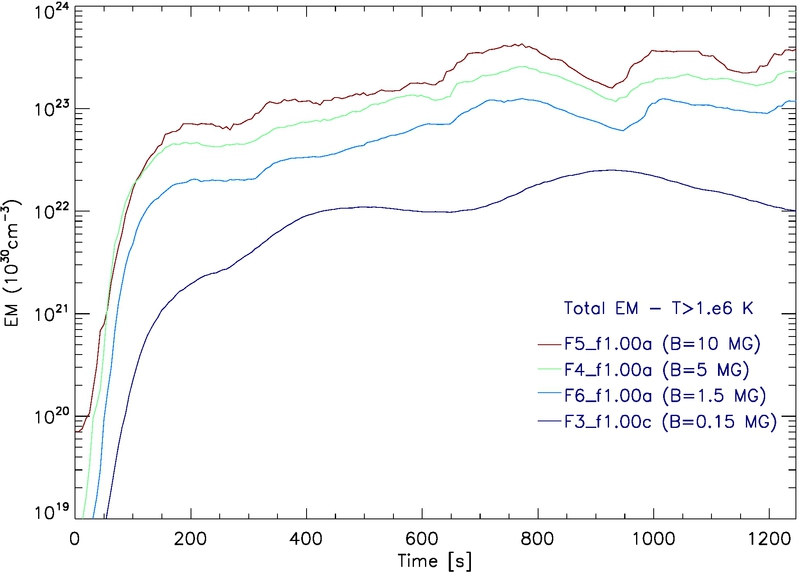}
	\caption{Total emission measure vs. time of plasma hotter than 1 MK. After few minutes, the value of emission measure approaches a virtually steady-state condition (run F4f1.00a).} 
	\label{fig:EM_gt1e6}
\end{figure}

Since we are interested in investigating the thermal X-ray emission originating from the activity occurring on the surface of the disk, finally, we analysed the emission measure (see Section \ref{sec:XraySynth}) of the plasma at temperature capable of producing thermal X-ray emission ($T>10^{6}K$). Fig. \ref{fig:EM_gt1e6} shows that, immediately after the beginning of the release of heat pulses, the amount of plasma at temperature greater than $1\;MK$, increases steadily reaching an almost steady-state condition after 100 seconds. The values of emission measure found indicate that the flaring activity leads to a strong thermal X-ray emission. 

\begin{figure*}
	\begin{overpic}[trim={0 1cm 0 1cm},clip,width=\columnwidth,tics=10]{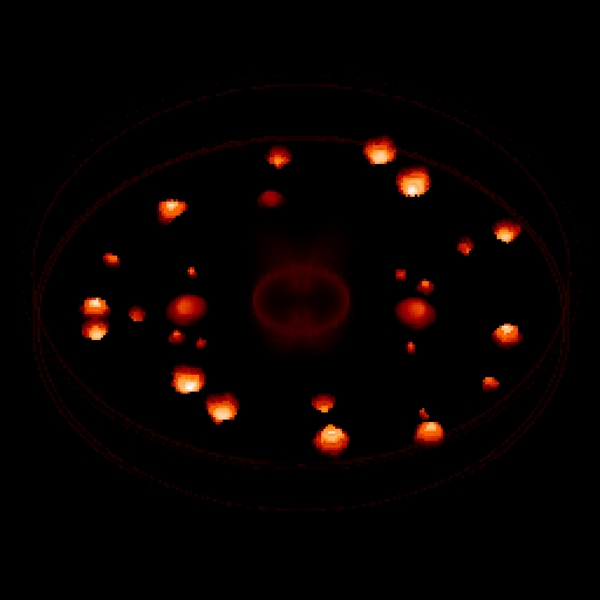}
		\put (1,56) {\textsf{\large\color{white}\textbf{Time: 62.3 s}}}
	\end{overpic}
	\includegraphics[trim={0 1cm 0 1cm},clip,width=\columnwidth]{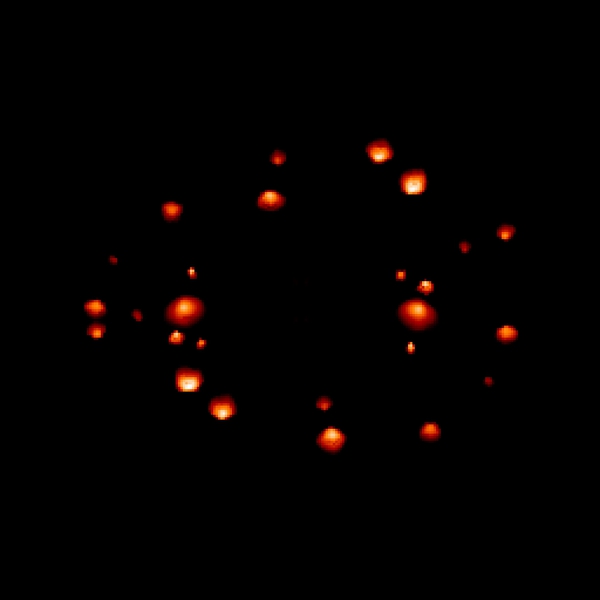}
	
	\vspace{1mm}		
	\begin{overpic}[trim={0 1cm 0 1cm},clip,width=\columnwidth,tics=10]{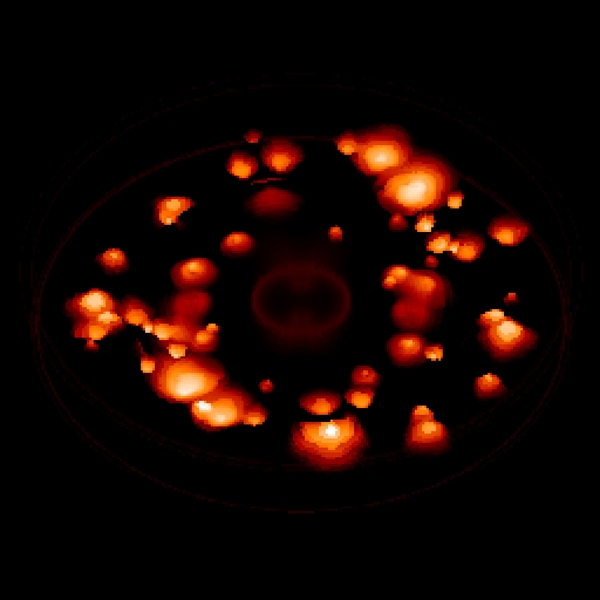}
		\put (1,56) {\textsf{\large\color{white}\textbf{Time: 124.6 s}}}
	\end{overpic}
	\includegraphics[trim={0 1cm 0 1cm0},clip,width=\columnwidth]{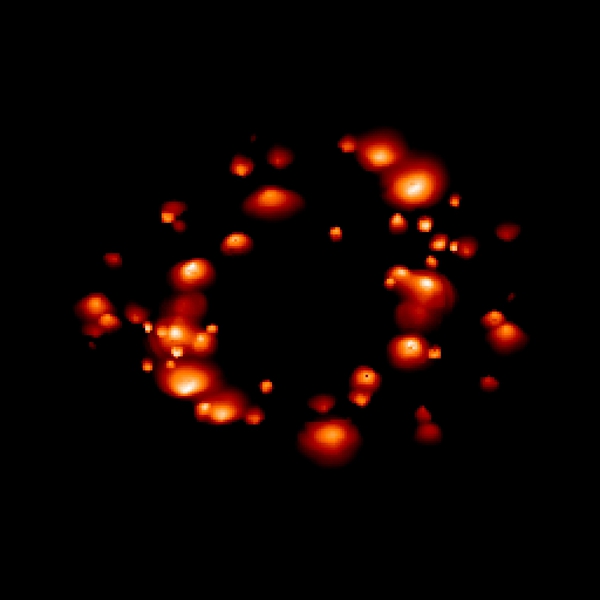}
	
	\vspace{1mm}
	\begin{overpic}[trim={0 1cm 0 1cm},clip,width=\columnwidth,tics=10]{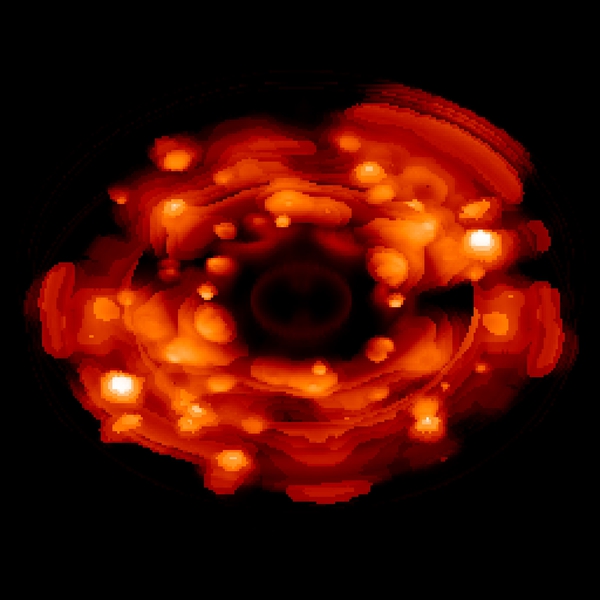}
		\put (1,56) {\textsf{\large\color{white}\textbf{Time: 311.5 s}}}
	\end{overpic}
	\includegraphics[trim={0 1cm 0 1cm},clip,width=\columnwidth]{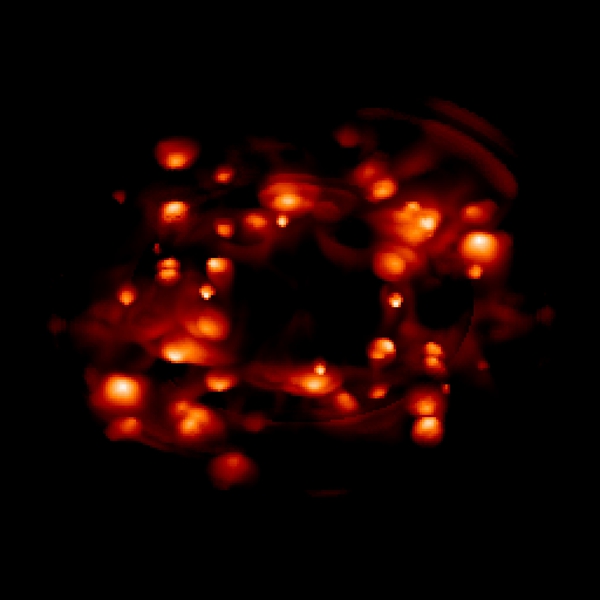}
	
	\vspace{1mm}
	\begin{overpic}[trim={0 1cm 0 1cm},clip,width=\columnwidth,tics=10]{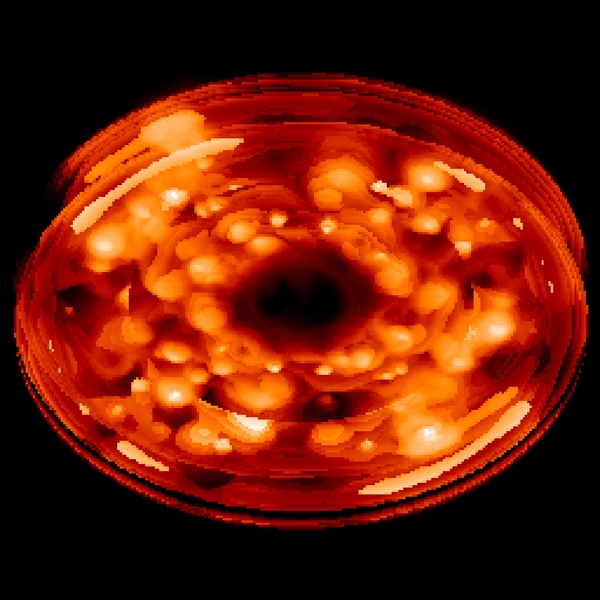}
		\put (1,56) {\textsf{\large\color{white}\textbf{Time: 623.0 s}}}
	\end{overpic}
	\includegraphics[trim={0 1cm 0 1cm},clip,width=\columnwidth]{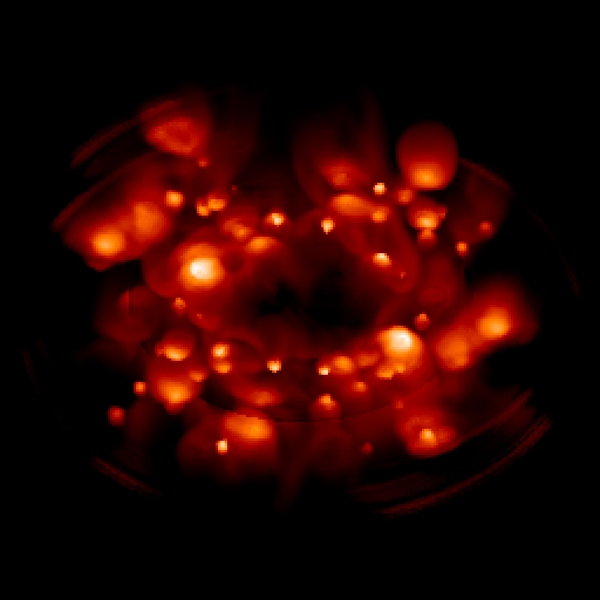}
	
	\vspace{1mm}
	\centering
	\includegraphics[trim={0 0 0 0},clip,width=0.7\textwidth]{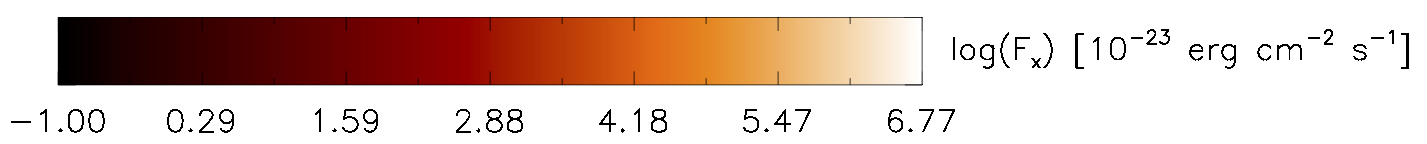}
	
	\caption{X-ray images of the whole domain as seen by an observer with a line of sight of 45$^{\circ}$ with the equatorial plane. On the left column panels soft X-ray $[0.1-2.0]\;keV$ flux values, on the right hand side panels hard X-ray $[2.0-10]\;keV$ flux values. The flux values are computed assuming a distance of 500 pc and no interstellar absorption. The images refers to the run F4f1.00a and are taken at the same times as those of Fig. \ref{fig:CoronalLoops}.} 
	\label{fig:Ximg}
\end{figure*}

From the model results, we synthesized the thermal X-ray emission originating from the whole system along the whole interval of time considered in our simulations. The method applied is described in Section \ref{sec:XraySynth}. The X-ray emission has been synthesized in two bands: the soft ($[0.1-2.0]\;keV$) and the hard ($[2.0-10]\;keV$). 
Fig. \ref{fig:Ximg} shows some of the synthesized X-ray images. Each image shows the simulated spatial distribution of X-ray flux observed at different times during the evolution of the system. 
The system is observed from a line of sight forming a 45$^{\circ}$ angle with the equatorial plane. 
From these X-ray images we can infer where the coronal emission is more conspicuous. We identify the feet of the coronal loops as the regions where the X-ray emission is concentrated. 
On the whole, the flaring activity seems to be capable of triggering the formation of a hot and tenuous layer of plasma extending on the disk surface. This layer is accountable for the bulk of the coronal X-ray emission, that is concentrated in proximity of the disk surface. 

\begin{figure*}[h!]
	\centering
	\includegraphics[trim={0 0 0.20cm 0.21cm},clip,width=9.1cm]
	{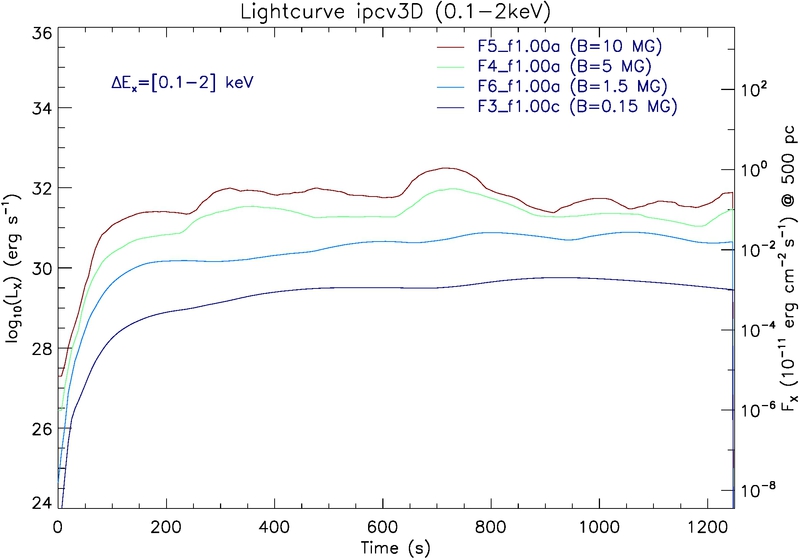}
	\includegraphics[trim={0.21cm 0 0 0.21cm},clip,width=9.1cm]
	{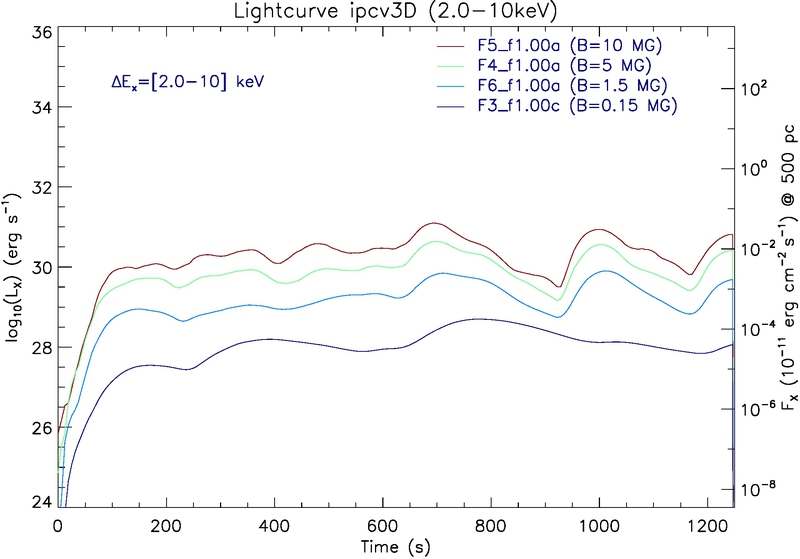}
	\caption{Comparative plot of lightcurves relative to various configurations explored for different values of magnetic field. On the left panel, soft X-ray band $[0.1-2.0]\;keV$ lightcurves; on the right panel, hard X-ray band $[2.0-10]\;keV$ lightcurves.} 
	\label{fig:LC_models}
\end{figure*}

\begin{figure*}[h!]
	\centering
	\includegraphics[trim={0 0 0.20cm 0.21cm},clip,width=9.1cm]
	{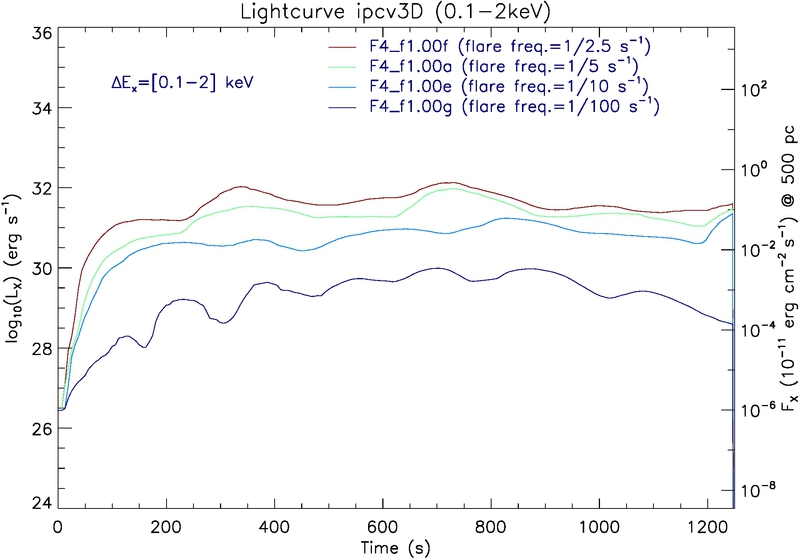}
	\includegraphics[trim={0.21cm 0 0 0.21cm},clip,width=9.1cm]
	{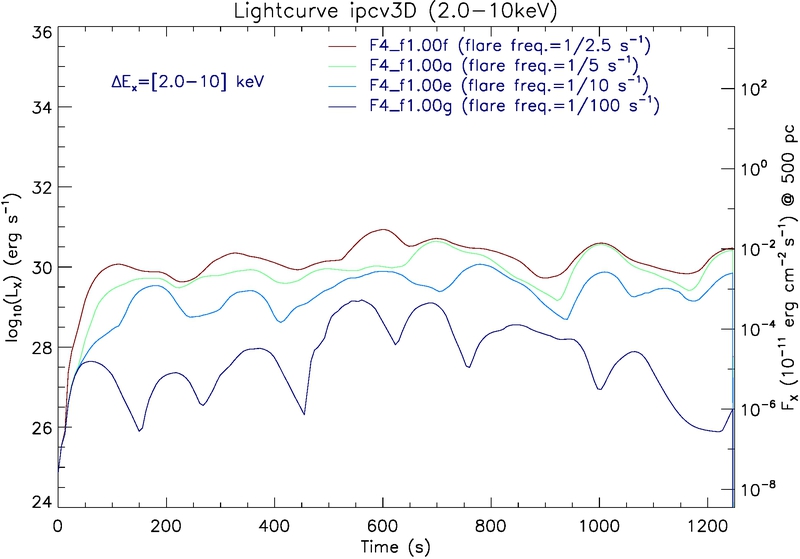}
	\caption{Comparative plot of lightcurves relative to the reference model. The explored parameter is the frequency of the energy pulse (flares). On the left panel, soft X-ray band $[0.1-2.0]\;keV$ lightcurves; on the right panel, hard X-ray band $[2.0-10]\;keV$ lightcurves.} 
	\label{fig:LC_frequency}
\end{figure*}

\begin{figure*}[h!]
	\centering
	\includegraphics[trim={0 0 0.20cm 0.21cm},clip,width=9.1cm]
	{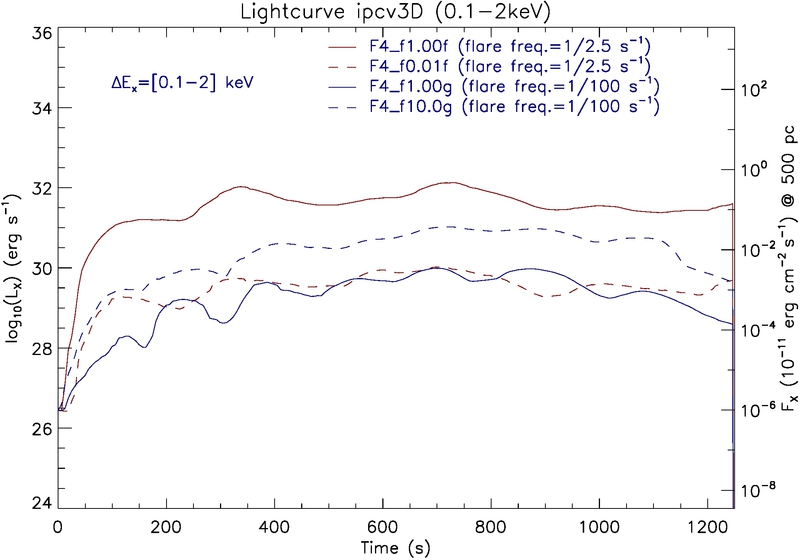}
	\includegraphics[trim={0.21cm 0 0 0.21cm},clip,width=9.1cm]
	{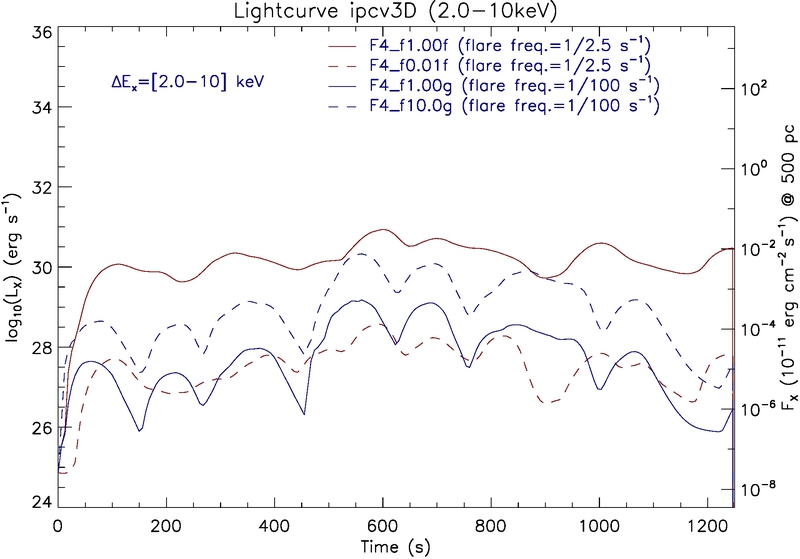}
	\caption{Comparative plot of lightcurves relative to the reference model. The explored parameters are the frequency and the intensity of the energy pulse (flares). On the left panel, soft X-ray band $[0.1-2.0]\;keV$ lightcurves; on the right panel, hard X-ray band $[2.0-10]\;keV$ lightcurves.} 
	\label{fig:LC_FvI}
\end{figure*}

The flux values of the thermal X-ray emission have been used in order to obtain the lightcurves for each model. The flux is defined as the total energy per unit of surface and per unit of time and the corresponding luminosity is obtained from the flux values supposed to be measured at a distance of 500 pc. 
Fig. \ref{fig:LC_models} shows the comparison between the lightcurves of the various configuration studied imposing a frequency of a flare every 5 seconds (assumed as reference value).
The luminosity, in every configuration explored reaches in few minutes the quasi-stationary condition and exhibits a clear oscillation. This effect is actually less noticeable in low field configurations (i.e. runs F3f1.00c and F6f1.00a) because these configurations are characterized by low density values. The density, through the term $n_{e}n_{H}$ (see Eq. (\ref{eq:EneCon})), strongly affects the radiative cooling. So, in low density configurations, the energy injected by each heat pulse is more slowly dissipated than in high density ones, and, when another pulse is released, the effect of the energy injected by the previous one has not been dissipated yet. In this case the effect of each flare is less evident in the global light curve.

Another key element, that we can infer from the lightcurves, is the maximum luminosity achieved: the denser is the disk (and, consequently, given the way our model is devised, the stronger the magnetic field) the higher is the luminosity. In fact, a denser plasma determines a higher value of emission measure, hence the contribution to the emission is greater. 
Comparing the lightcurves of the harder band with the corresponding ones of the softer band, we find a hardness ratio of the spectra, namely the ratio between hard and soft X-ray luminosity, approximately $\lesssim0.1$.

In the context of the reference model we also explored the effect of the frequency and intensity of flares on the lightcurves. 
Fig. \ref{fig:LC_frequency} shows how the lightcurves are affected by different flare release frequencies. 
The difference between maximum and mean value ranges between a factor of 5 and 8 of the mean value in high flare frequency configurations (run F4f1.00f, F4f1.00a and F4f10.00e), whilst, in low flare frequency ones (run F4f1.00g and F4f10.00g), this value rises up to 57 times the mean value.
From this comparison, we can infer that the variability of X-ray emission strongly depends on the rate of triggered flares.
Actually, as long as the period between two subsequent flares is shorter than the duration of each energy pulse $(100 s)$, the variability of the luminosity value is less noticeable and the effect of each single flare is mixed and superimposed to the ones of the other neighbour flares.
We actually observe a background emission due to many small flares evolving simultaneously while higher peaks are due to the strongest flares.
When the period between flares is the same of the duration of the energy pulse, the effect of each single pulse is easily identifiable. We can observe the fast rise and the subsequent slow decay of the thermal X-ray emission characteristic of each flare. 
Indeed, for high temperature plasma the conductive and the radiative cooling are more effective (see Fig. \ref{fig:RadCool_cooltable}), causing a quicker decrease of emission.
This feature determines a noticeable modulation of luminosity as a sequence of partially overlapping bursts. 
Furthermore, the variability of the luminosity is generally more evident in the hard than in the soft band, because the hotter plasma is the dominant source of the hard emission.

Fig. \ref{fig:LC_FvI} shows a comparison between lightcurves, taken in the context of the reference model with the same flare frequency but different maximum intensity. This comparison shows that the X-ray variability strongly depends on the flare frequency, while the main effect of a higher intensity of flares is to bring up the average luminosity keeping, at the same time, the modulated shape almost unchanged. 

Finally, we note that modulation in X-ray/UV/Vis lightcurves may be observed in \emph{IPCV}s, due to the formation of a lighthouse beam, when the WD’s spin and magnetic axes are not aligned, and the star magnetic field rotates with a tilted axis.
This is due to the energy release of shocks generated by the accreting material impacting onto the star’s surface, creating an intense bright spot at one or both WD’s magnetic poles.
Hence, pulsations and additional periods can emerge when the rotating searchlight illuminates the various structures in the binary 
\citep[e.g.][]{Patterson1994,Romanova2003,Kulkarni&Romanova2008,Giovannelli2012,Romanova2013}.
Nevertheless, no pulsation of the coronal radiation has been observed so far \citep[e.g.][]{VenterMeintjes2007}. 
In our work, the seed magnetic field in the disk is due to the presence of a magnetized central object. 
In particular we adopted the simplest magnetosphere configuration: a non-tilted dipole. 
Hence, our model cannot reproduce any rotational modulation of the coronal radiation.  
Although the search for a possible pulsation of the coronal radiation is out of the scope of the present paper, we note that the lack of observed modulation may be closely related to the fact that flares are not dependent on the large-scale configuration of the magnetic field (in fact the formation of an extended corona does not require a central magnetized object) and strongly depend on the local magnetic field configuration in the disk.

\section{Conclusions}
\label{sec:Conclusions}

We investigated the effect of an intense flaring activity occurring on the surface of the accretion disk of various \emph{IPCV} prototypes. The prototypes taken into account consider different values of density and magnetic field.
We focused on different aspects: the formation of an extended corona above the disk; the contribution of the related coronal activity to the thermal X-ray emission from these objects; the conditions able to jeopardize the stability of the disk.

We have developed a 3D MHD model taking into account all the key physical processes: viscosity of the disk, magnetic field oriented thermal conduction, radiative cooling and gravitational forces.
In order to set up the initial conditions of the 3D simulations, we have devised a two-step technique. As a first step, we have run 2.5D simulations in order to obtain a developed configuration of the magnetic field.
From this we have reconstructed a 3D initial condition where we then simulated a hail of flares distributed stochastically on the surface of the inner part of the accretion disk. Finally, we synthesized the thermal X-ray emission from each system all along the whole evolution. 

In the light of our findings, we can draw the following main conclusions: 
(a) an intense flaring activity close to the surface of the accretion disk can build up an extended corona linking the star to the disk along the magnetic field lines; 
(b) the thermal X-ray emission from the corona is mainly emitted from a thin layer immediately above the surface of the disk;
(c) the thermal X-ray luminosity due to the flaring activity may contribute to the emission observed from these objects;
(d) this coronal activity can influence the disk configuration and its stability, effectively deforming the magnetic field lines, with effects strongly dependent on density and magnetic field intensity; 
(e) the radiative and conductive cooling has a crucial role in the modulation of luminosity due to the chain of flares, and it has different impact on the lightcurves depending on the X-ray band considered and on the temperature of the emitting plasma, contributing efficaciously to the phenomenon of flickering originating from the inner region of the disk.

It is worth noting that the synthesized thermal X-ray emission exhibits luminosity values ranging between $10^{30}$ and $10^{32}\;erg/s$. These values, especially those for the high density configurations explored, are in general agreement with previous studies \citep{Patterson1980,VanDerWoerd1987,EracleousHalpernPatterson1991,ChoiDotaniAgrawal1999,VenterMeintjes2007}

Summarizing, the results presented above suggest that an intense flaring activity occurring on the surface of the accretion disk of \emph{IPCV} stars may turn out to have an important effects on the structure of the system and its stability, and a key role in the formation of an extended corona, whose effects may contribute to the thermal X-ray emission observed from these objects.

\begin{acknowledgements}
	We wish to thank an anonymous referee for criticisms and helpful suggestions which have improved the paper. 
	This work was supported in part by the Italian Ministry of University and Research (MIUR) and by Istituto Nazionale di Astrofisica (INAF). PLUTO is developed at the Turin Astronomical Observatory in collaboration with the Department of Physics of Turin University. We acknowledge also the CINECA Award HP10CSHHKL and the HPC facility (SCAN) of the INAF-Osservatorio Astronomico di Palermo, for the availability of high performance computing resources and support.
\end{acknowledgements}

\bibliographystyle{aa}
\bibliography{bibliography}

\end{document}